\documentclass[a4paper,12pt]{article}
\usepackage{amsmath}
\usepackage{amssymb}
\usepackage{braket}
\usepackage{cancel}
\usepackage{graphicx}
\usepackage{color}
\usepackage[utf8]{inputenc}
\usepackage{hyperref}
\usepackage{multirow}
\usepackage{slashed}
\usepackage[normalem]{ulem}
\usepackage{wasysym}
\usepackage{amssymb}
\usepackage{braket}
\usepackage{simplewick}
\usepackage{array}
\usepackage{graphicx}
\usepackage{float}
\usepackage{titlesec}
\usepackage{listings}
\usepackage{hyperref}
\usepackage[english]{babel}
\usepackage{xcolor}
\usepackage{subcaption}

\newcommand{\mubox}{[\mu]}
\newcommand{\taubox}{[\tau]}

\newcommand{\barnumubox}{[\bar{\nu}_\mu]}
\newcommand{\nutaubox}{[\nu_\tau]}
\newcommand{\barnutaubox}{[\bar{\nu}_\tau]}
\newcommand{\nunubox}{[\nu_{\tau} \bar{\nu}_\mu]}

\newcommand{\mtau}{m_\tau}

\usepackage[english]{babel}

\usepackage{graphicx}

\usepackage{geometry}
\begin{document}

\begin{titlepage}

\begin{flushright}
{\small
P3H-21-027 \\ 
SI-HEP-2021-13 \\
Nikhef-2021-011 \\
\today \\
}
\end{flushright}

\vskip1cm
\begin{center}
{\Large \bf\boldmath 
Impact of background effects on the \\[2mm] inclusive $V_{cb}$ determination }
\end{center}

\vspace{0.5cm}
\begin{center}
{\sc Thomas Mannel,$^a$ Muslem Rahimi,$^a$ 
and K. Keri Vos$^{b,c}$} \\[6mm]
{\it $^a$ Center for Particle Physics Siegen (CPPS), \\
Theoretische Physik 1, Universit{\"a}t Siegen, \\
57068 Siegen, Germany\\[0.3cm]

{\it $^b$Gravitational 
Waves and Fundamental Physics (GWFP),\\ 
Maastricht University, Duboisdomein 30,\\ 
NL-6229 GT Maastricht, the
Netherlands}\\[0.3cm]

{\it $^c$Nikhef, Science Park 105,\\ 
NL-1098 XG Amsterdam, the Netherlands}}
\end{center}

\vspace{0.6cm}
\begin{abstract}
\vskip0.2cm\noindent
The determination of the CKM element $V_{cb}$ from inclusive semileptonic $b\to c \ell \bar\nu$ decays has reached a high precision thanks to a combination of theoretical and experimental efforts. Aiming towards even higher precision, we discuss two processes that contaminate the inclusive $V_{cb}$ determination; the $b\to u$ background and the contribution of the tauonic mode: $b\to c(\tau \to \mu\nu\bar{\nu})\bar{\nu}$. Both of these contributions are dealt with at the experimental side, using Monte-Carlo methods and momentum cuts. However, these contributions can be calculated with high precision within the Heavy-Quark Expansion. In this note, we calculate the theoretical predictions for these two processes. The $b\to u$ results are compared with generator-level Monte-Carlo results used at Belle and Belle II. We have good agreement between theory and Monte-Carlo for lepton energy moments, but less for hadronic mass moments. Based on our results the uncertainties due to these backgrounds processes can basically be eliminated by properly including them into the analyses.
\end{abstract}
\end{titlepage}

\section{Introduction}
\label{sec::Introduction}
The determination of $V_{cb}$ from inclusive $b\to c \ell \nu$ decays relies on the heavy quark expansion (HQE)
and is in a very mature state (see e.g.~\cite{Manohar:2000dt, Grozin:2004yc, Mannel:2010wj, Mannel:2019qel}). The current inclusive determination of $V_{cb} =(42.21\pm0.78)\cdot 10^{-3}$ \cite{Gambino:2013rza,Alberti:2014yda} has an impressive $2\%$ uncertainty. The analysis includes perturbative corrections up to $\alpha_s^2$ terms \cite{Jezabek:1988iv,Pak:2008cp,Melnikov:2008qs,Biswas:2009rb,Gambino:2011cq} and power corrections up to $1/m_b^3$. Higher-order terms up to $1/m_b^5$ have been classified \cite{Dassinger:2006md} and studied using a lowest-lying state approximation \cite{Gambino:2016jkc}. These higher-order terms are currently studied in more detail \cite{WorkInProgress1} using a new method to determine inclusive $V_{cb}$ from $q^2$-moments \cite{Fael:2018vsp} using reparametrization invariance \cite{Mannel:2018mqv}. Recently, also $\alpha_s^3$ corrections \cite{Fael:2020tow} and $\alpha_s$ corrections to $1/m_b^3$ terms \cite{Mannel:2019qel} were studied. In combination with the expected
data from Belle II, we therefore foresee a very precise determination of $V_{cb}$.

The current tension between the value of $V_{cb}$ obtained from the
inclusive determination and the one obtained from the exclusive channels $B \to D \ell \bar{\nu}$
and $B \to D^* \ell \bar{\nu}$ indicates that there may still be systematic effects which need to be understood better (see e.g. \cite{Bordone:2019vic, Bordone:2019guc, Gambino:2019sif, Bigi:2017jbd}).
While the exclusive determination requires the input of form factors which are taken from lattice
simulations (and/or using QCD sum rules), the inclusive determination is assumed to be theoretically cleaner, as the required hadronic matrix elements can be obtained from data, at least up to the order $1/m_b^3$ and
possibly even to $1/m_b^4$. Therefore, the inclusive $V_{cb}$ determination may be pushed towards even higher precision. 

In this quest even tiny effects and backgrounds need to be carefully studied before a precision at the level of one percent (or even less) can be claimed. The current method for extracting $V_{cb}$ relies on taking lepton energy and hadronic invariant mass moments of the  $B \to X_c \ell \bar{\nu}$ decay where $\ell = \mu, e$. However, the data taken at the $B$ factories are based on the inclusive
$B \to X \ell$ rate, from which the $B \to X_c \ell \bar{\nu}$ is extracted using Monte-Carlo simulations. While the $B \to X_c \ell \bar{\nu}$ is certainly the dominant part, aiming towards a sub-percent precision requires a modified approach. Overall, we identify two processes that contaminate the $B\to X_c$ signal and that could be constrained using the HQE:

\begin{itemize}
            \item {\boldmath \bf The contribution of $b \to u \ell \bar{\nu}$:} Although this contribution is suppressed by a factor $(V_{ub}/V_{cb})^2$ and thus is not expected
            to make a significant contribution, extreme precision will require to have a good control of it.
            \item {\bf \boldmath The contribution of $b \to c (\tau \to \ell \bar{\nu} \nu) \bar{\nu}$:} This
            contribution is suppressed only by the smaller phases space and by the branching fraction
            of $\tau \to  \ell \nu \bar{\nu}$. It may be reduced by appropriate cuts, but still needs to be described with the corresponding
            precision.
\end{itemize}             
In the current note, we address these two contributions. Within the HQE, total rates as well as various moments of kinematic distributions can be reliably
calculated and compared to the data. This may also include QED corrections, since in $B \to X \ell$
the $X$ may not only include neutrinos but also photons. Obviously the HQE result for $B \to X \ell$
will depend not only on $m_b$, $m_c$ and $V_{cb}$ and the HQE parameters, but also on $V_{ub}$ and
$m_\tau$.
Using HQE calculations, we may check the quality of the extraction of $B \to X_c \ell \bar{\nu}$
from $B \to X \ell$ by comparing our HQE results with the Monte-Carlo simulations of the $b \to u \ell \bar{\nu}$. In this note
we study this comparison. For the $b \to c (\tau \to \ell \nu \bar{\nu}) \bar{\nu}$, this comparison is more cumbersome due to different experimental cuts and Monte-Carlo data is at the moment not available. For this five-body decay, we therefore only provide the theoretical predictions. To our knowledge, no theoretical study of these effects in inclusive decays has been performed.

For the upcoming Belle II analyses we suggest to change the strategy for the extraction of $V_{cb}$
by comparing the data on $B \to X \ell$ directly to the corresponding theoretical expressions,
circumventing the problem of constructing first the data for $B \to X_c \ell \bar{\nu}$ by Monte-Carlo procedures.
The aim of this note is to facilitate this strategy by supplying the necessary theoretical expressions. In Section \ref{ch::Inclusive-three-body} we discuss the lepton energy, hadronic invariant mass and $q^2$-moments for the three-body decay $b \to u \ell \nu$. We compare our result with the Monte-Carlo results. Moreover, we discuss the aforementioned moments for the five-body decay $b \to c (\tau \to \ell \nu \bar{\nu}) \bar{\nu}$ in Section \ref{ch::Inclusive-five-body}. We compare our results with the moments of three-body $b \to c \ell \nu$ decay. Finally, we conclude in Section \ref{ch::Conclusion}.

\section{Background from the $\bar{B} \to X_{u} \ell \bar{\nu}_{\ell}$ decay}
\label{ch::Inclusive-three-body}
The $e^+ e^-$ $B$ factories Belle (II) and BaBar have a very clean environment, 
such that a fully inclusive measurement of $B \to X \ell$ can be performed, which is the basis of the current inclusive $V_{cb}$ determination. In the current analyses the contribution of the $b \to c \ell \bar{\nu}$ transition is extracted from $B \to X \ell$ using Monte Carlo simulations of the backgrounds. After the subtraction of the backgrounds the resulting data is compared to the theoretical predictions for $B \to X_c \ell\bar{\nu}$. This procedure induces uncertainties related to the Monte Carlo simulations. On the other hand, the fully inclusive $B \to X \ell$ can be predicted theoretically at the same level of precision as $B \to X_c \ell \bar{\nu}$, so the process of background subtraction can be avoided completely, or at least to a large extend. To this end, one may 
compute the inclusive rate for $B \to X \ell$ by adding the contributions
\begin{align}
\text{d} \Gamma (B \to X \ell) = \text{d} \Gamma (B \to X_c \ell \bar{\nu}) + \text{d} \Gamma (B \to X_u \ell \bar{\nu}) + 
\text{d} \Gamma (B \to X_c (\tau \to \ell \bar{\nu} \nu) \bar{\nu})
\end{align}
For fully inclusive observables such as (cut) moments each term on the right-hand side
can be computed individually in terms of the standard HQE based on the local OPE.  

In the following we will use the known results of the HQE for $\text{d} \Gamma (B \to X_u \ell \bar{\nu})$ in the local OPE to compare to generator-level Monte Carlo results, which are used for the background subtraction. In the next section, we compute the five-body contribution $\text{d} \Gamma (B \to X_c (\tau \to \ell \bar{\nu} \nu) \bar{\nu})$.

\subsection{Set-up for inclusive decays}
The semileptonic $b\to u \ell \bar\nu$ decay is described by 
\begin{align}
    \mathcal{H}_{W} &= \frac{G_{F} V_{ub}}{\sqrt{2}} J_{L}^{\alpha} J_{H, \alpha} + \, \, \text{h.c.},
\end{align}
where $J_{L}^{\alpha} = \bar{\ell} \gamma^{\alpha}(1-\gamma_5) \nu$ and $J_{H, \alpha} = \bar{u} \gamma_\alpha (1- \gamma_5) b$ are the leptonic and hadronic currents, respectively. Equivalent as for the $\bar{B}\to X_c \ell \bar{\nu}_\ell$, we obtain the triple differential decay rate: 
\begin{align}
    \frac{\text{d}\Gamma}{\text{d}E_{\ell} \text{d}q^2 \text{d}E_{\nu}} &= \frac{G_{F}^2 |V_{ub}|^2} {16 \pi^3} L_{\mu \nu} W^{\mu \nu} ,
    \label{eq::LWcontraction}
\end{align}
where $E_{\ell (\nu)}$ is the lepton (neutrino) energy and $q^2$ is the dilepton invariant mass. 
Here $L_{\mu \nu}$ is the leptonic tensor and $W^{\mu\nu}$ is the hadronic tensor:
\begin{align}
    W^{\mu \nu} &= \frac{1}{4} \sum_{X_u} \frac{1}{2 m_{B}} (2\pi)^3 \bra{\bar{B}}J_{H}^{\dagger \mu} \ket{X_{u}} \bra{X_{u}} J_{H}^{\nu} \ket{\bar{B}} \delta^{(4)}(p_{B} -q - p_{X_{u}}) \ , 
    \label{eq::HadronTensor}
\end{align}
where $p_{X_u}$ is the total partonic momentum. 
Decomposing (\ref{eq::HadronTensor}) into Lorentz scalars gives
\begin{align}
    W^{\mu \nu} &= -g ^{\mu \nu} W_{1} + v^{\mu} v^\nu W_{2} - i \epsilon^{\mu \nu \rho \sigma} v_{\rho} q_{\sigma} W_3 + q^\mu q^\nu W_4 + (q^\mu v^\nu + v^\mu q^\nu) W_5.
    \label{eq::HadronDecompose}
\end{align}
leading to 
\begin{align}
\label{eq:tripdif}
    \frac{\text{d}\Gamma}{\text{d}E_{\ell} \text{d}q^2 \text{d}E_{\nu}} &= \frac{G_{F}^2 |V_{ub}|^2}{2 \pi^3} \left[q^2 W_{1} + (2 E_{\ell} E_\nu - \frac{q^2}{2}) W_2 + q^2 (E_{\ell} - E_{\nu}) W_3 \right. \nonumber \\
    & \left. \frac{1}{2} m_{\ell}^2 \left( -2 W_{1} + W_2 -2 (E_\nu+ E_\ell)W_{3} + q^2 W_4 + 4 E_\nu W_5 \right) - \frac{1}{2} m_{\ell}^4 W_4 \right] \ ,
\end{align}
where we have omitted explicit $\theta$-functions. When considering $\ell = e, \mu$, we set $m_\ell \to 0$, such that $W_{4,5}$  do not contribute. The $W_{1,2,3}$ are now obtained using the HQE.

The HQE has become an well-established tool in the study of inclusive $B$ meson decays, allowing the expression of observables in a double expansion of $\alpha_s$ and $1/m_b$. It is set up by redefining the heavy-quark field by splitting the momentum $p_Q$ of the heavy quark as $p_Q = m_Q v + k$, where $v$ is a time-like vector and $k$ the residual momentum. We can expand the residual momentum $k \sim \mathcal{O}(\Lambda_{\text{QCD}})$ which yields the standard operator-product expansion (OPE), which separates the short-distance physics from non-perturbative forward matrix elements which contain chains of covariant derivatives (see e.g. \cite{Manohar:2000dt}). This introduces the hadronic matrix elements, $\mu^2_G$ and $\mu^2_\pi$ at $1/m_b^2$ and $\rho^3_D$ and $\rho^3_{\rm LS}$ at $1/m_b^3$ which are defined as

\begin{align}
    2 m_{B} \mu_{\pi}^2 &= - \bra{B(v)} \bar{b}_{v}  (i D)^2 b_{v} \ket{B(v)} , \\
    2 m_{B} \mu_{G}^2 &=  \bra{B(v)} \bar{b}_{v}  (i D_\mu) (i D_\nu) (-i \sigma^{\mu \nu}) b_{v} \ket{B(v)} , \\
    2 m_{B} \rho_{D}^3 &=  \bra{B(v)} \bar{b}_{v}  (i D_\mu) (i v \cdot D) (i D^{\mu}) b_{v} \ket{B(v)} , \\
    2 m_{B} \rho_{LS}^3 &=  \bra{B(v)} \bar{b}_{v}  (i D_\mu) (i v \cdot D) (i D_\nu) (-i \sigma^{\mu \nu}) b_{v} \ket{B(v)}.
\end{align}
(and a proliferation of matrix elements at higher orders \cite{Dassinger:2006md,Mannel:2018mqv, Fael:2018vsp}).

In the following, we use this local OPE to compute different moments of the $b\to u \ell \bar\nu$ spectrum. The procedure follows closely the standard derivation of the moments in $b\to c \ell \bar\nu$.

\subsection{Definition of the moments}
In order to obtain the $b\to u$ local contribution, we calculate different moments of the spectrum. We define normalized moments for a given observable denoted as $\mathcal{O}$:
\begin{align}\label{eq:Oexpr}
    \braket{\mathcal{O}^{n}}_{E_\ell > E_\ell^{\text{cut}}} &= \frac{\int_{E_\ell > E_\ell^{\text{cut}}} \text{d}\mathcal{O} \, \mathcal{O}^n \frac{\text{d} \Gamma}{\text{d} \mathcal{O}} } {\int_{E_\ell > E_\ell^{\text{cut}}} \text{d}\mathcal{O} \, \frac{\text{d} \Gamma}{\text{d} \mathcal{O}} },
\end{align}
where $E_\ell^{\text{cut}}$ is the energy cut of the lepton $\ell = (e, \, \mu)$ and $n$ denotes the $n$-th order of moment. In addition, we define central moments:
\begin{align}
\braket{ (\mathcal{O} -  \braket{\mathcal{O}})^{n}} &= \sum_{i=0}^{n} \, \left(\begin{array}{c} n \\ i  \end{array}\right)\braket{(\mathcal{O})^i} (- \braket{\mathcal{O}})^{n-i} \, , \hspace{0.5cm}
\, .
\end{align}

\begin{table}[t]
\begin{center}
\begin{tabular}{ c | c }
\hline\hline
$m^{\text{kin}}_{b}$ & (4.546 $\pm$ 0.021) GeV 
\\
\hline
$\overline{m}_{c}$(3 GeV) & (0.987 $\pm$ 0.013) GeV 
\\ 

\hline
$\left(\mu_{\pi}^{2}(\mu)\right)_{\text{kin}}$ & (0.432 $\pm$ 0.068) GeV$^{2}$ 
\\
\hline
$\left(\mu_{G}^{2}(\mu)\right)_{\text{kin}}$ & (0.360 $\pm$ 0.060) GeV$^{2}$ 
\\
\hline
$\left(\rho_{D}^{3}(\mu)\right)_{\text{kin}}$ & (0.145 $\pm$ 0.061) GeV$^{3}$
\\
\hline
$\left(\rho_{LS}^{3}(\mu)\right)_{\text{kin}}$ & (-0.169 $\pm$ 0.097) GeV$^{3}$ 
\\
\hline
$\alpha_{s}(m_{b})$ & 0.223 
\\
\hline\hline
\end{tabular}
\end{center}
\caption{Numerical inputs taken from \cite{Gambino:2016jkc}. For the charm mass, we use the $\overline{\text{MS}}$ scheme at 3 GeV. All other hadronic parameters are in the kinetic scheme at $\mu = 1$ GeV.}
\label{table::input}
\end{table}
Specifically, we discuss the lepton energy moments $ \braket{E_\ell^{n}}$, hadronic mass moments $\braket{M_{x}^{n}}$ and $q^2$ ($q^2 = (p_\ell + p_\nu)^2$) moments $\braket{\left(q^{2}\right)^n}$. The moments can be obtained using Eq.~\eqref{eq:Oexpr} and the triple differential rate in Eq.~\eqref{eq:tripdif}. For the hadronic invariant mass, this requires its relation to the partonic quantities: 
\begin{align}
\braket{M_{x}^{2}} &=  \braket{p^2_{X_{u}}} + \bar{\Lambda} m_{b} \braket{z} + \bar{\Lambda}^2 \label{eq::Mx1}
\end{align}
where $z \equiv 2(v\cdot p_{X_u})/m_b$ and 
\begin{align}
    \bar{\Lambda} &= m_{B} -m_{b} - \frac{\mu_\pi^2-\mu_G^2}{2 m_b} + \ldots \ ,
    \label{eq::Lambda_bar}
\end{align}
where the ellipses denote terms of higher orders in the $1/m_{b}$ expansion.
 In our HQE calculation, we include power-corrections up to order $\mathcal{O}(1/m_{b}^3)$ and radiative-corrections of order $\mathcal{O}(\alpha_s)$ to the partonic expression  (see also \cite{Gambino:2005tp}). The numerical input parameters for the computation of the moments are taken from \cite{Gambino:2016jkc} and given in Table \ref{table::input}. In order to avoid renormalon ambiguities related to the pole mass, we work in the short-distance kinetic mass scheme (used in \cite{Gambino:2016jkc} to extract $V_{cb}$). We can relate the pole scheme to the kinetic scheme through a perturbative series, see Appendix \ref{sec::mass-scheme}.

\subsection{Comparison between theory and Monte-Carlo}
\label{sec::comptheoexp}

In order to discuss the reliability of the background-subtraction procedure based on 
MC simulations we perform a direct comparison of the MC data with the theoretical prediction 
for the moments. To this end, we compare the moments extracted from MC simulations for 
the $b \to u$ transition only with the theoretically predicted moments, including again only 
the $b \to u$ transitions.

\begin{figure}[h]
	\centering
	\subfloat{\includegraphics[width=0.5\textwidth]{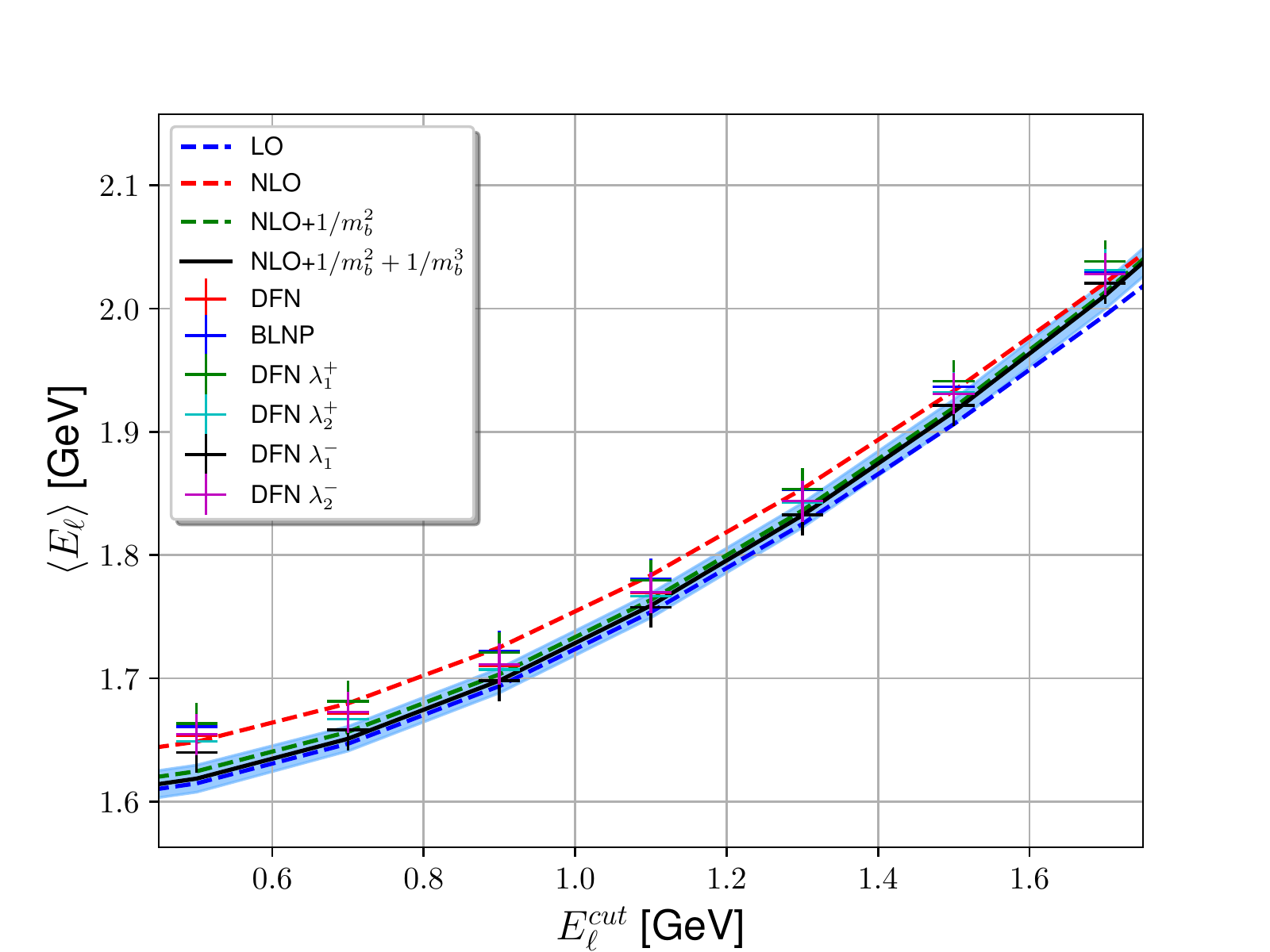}} 
  	\subfloat{\includegraphics[width=0.5\textwidth]{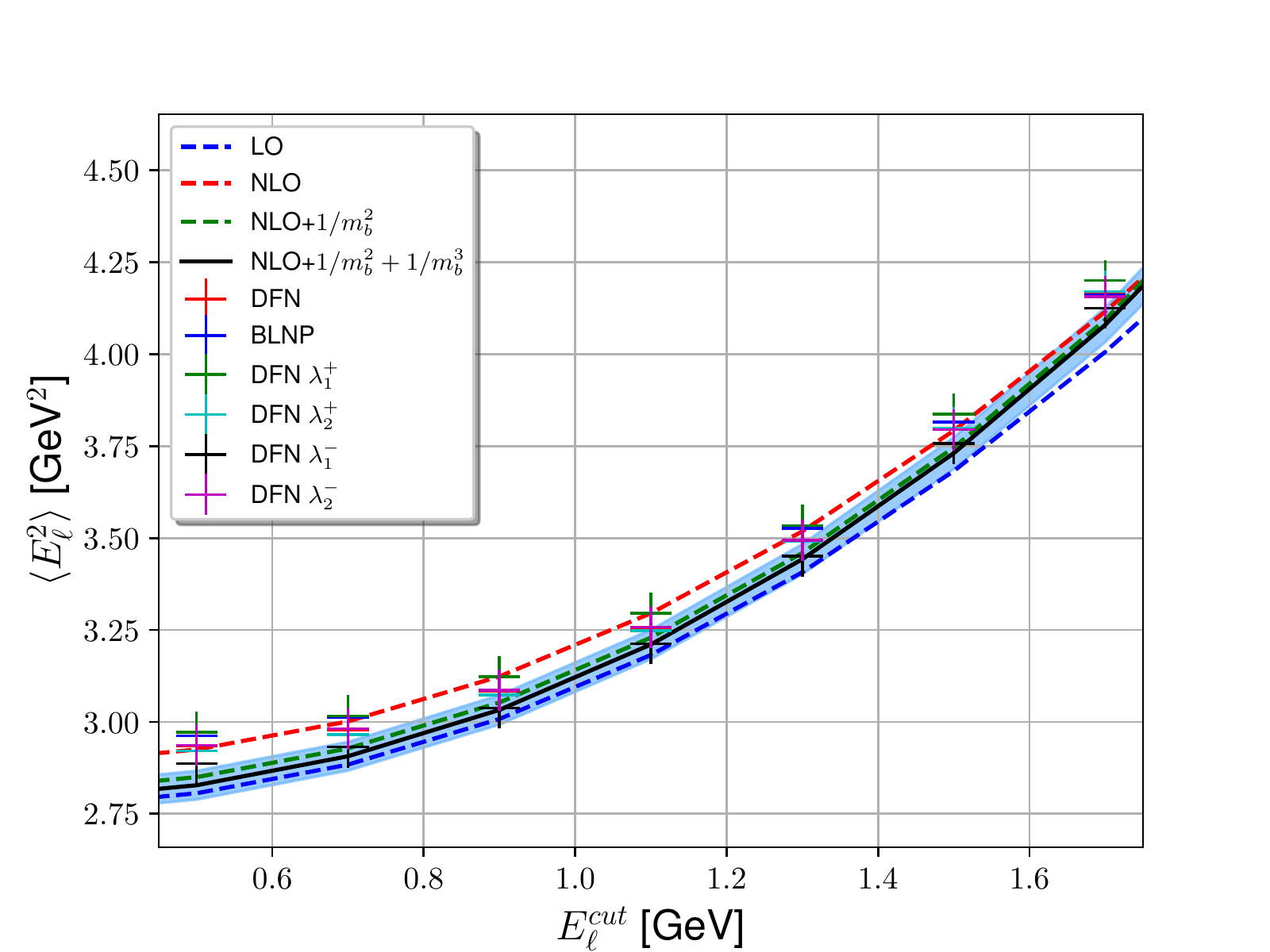}} \\
  	\subfloat{\includegraphics[width=0.5\textwidth]{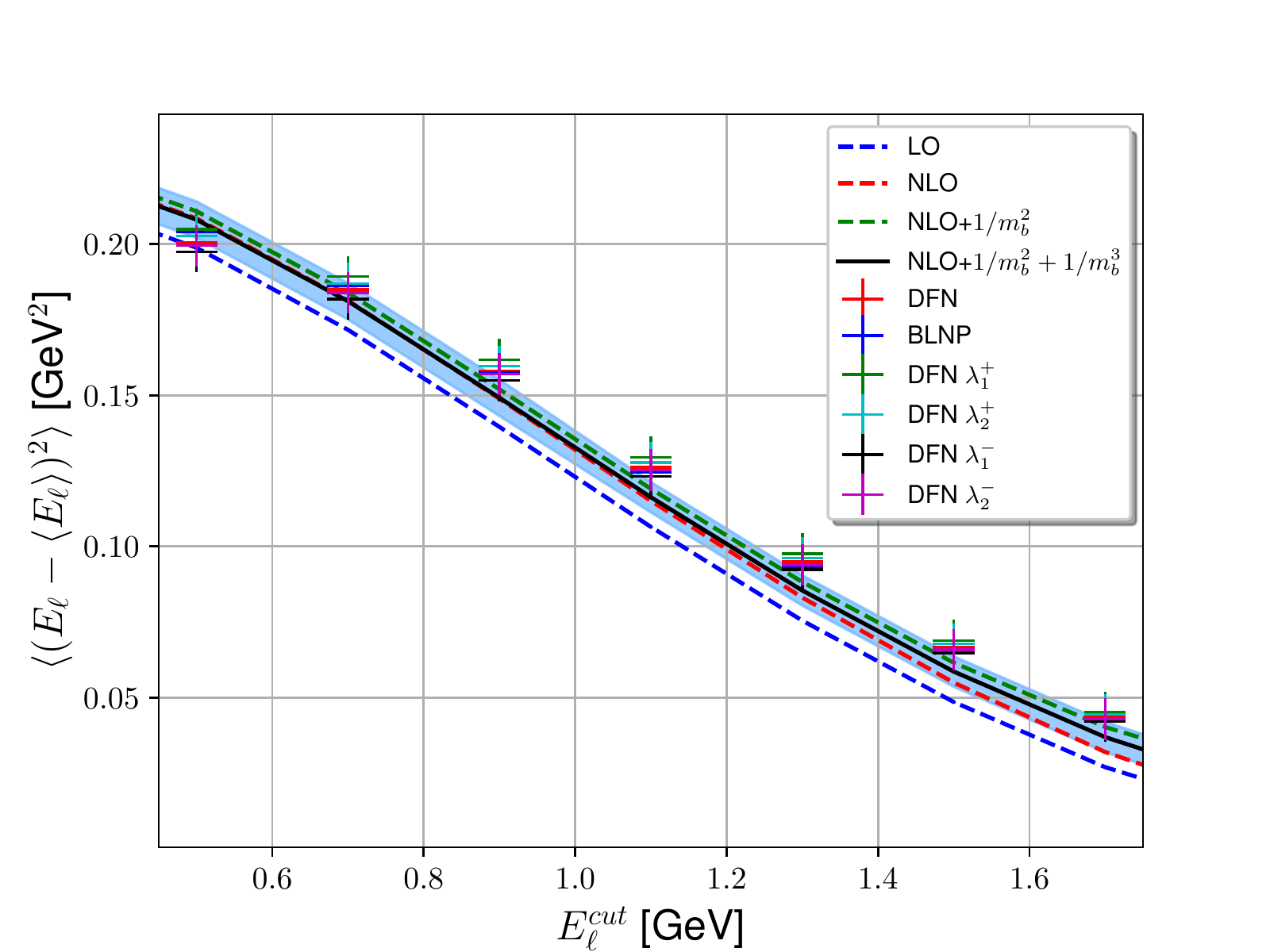}}
  	\subfloat{\includegraphics[width=0.5\textwidth]{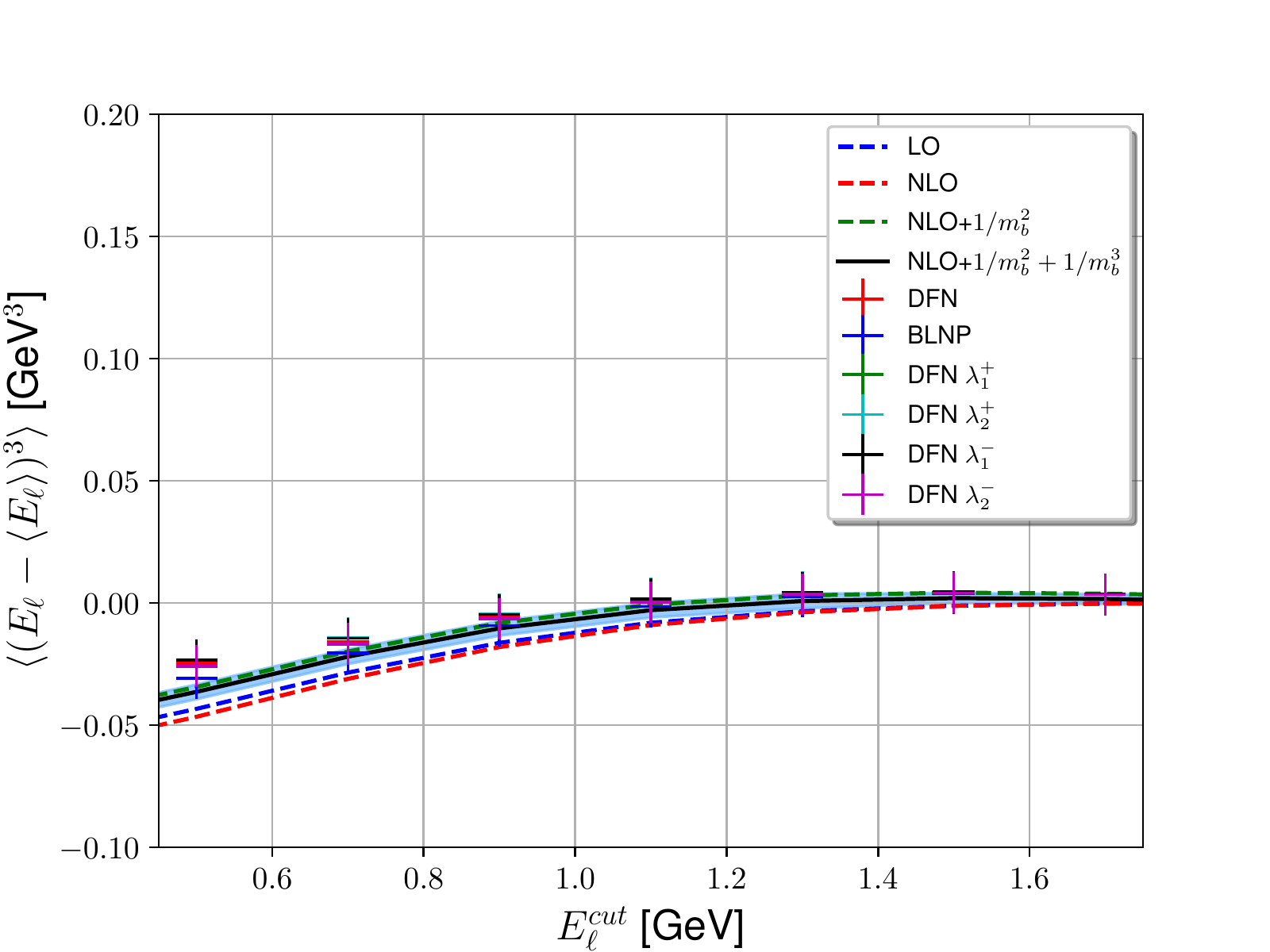}} 
  \caption{HQE results for lepton-energy moments with different energy cuts, showing leading order (LO), next-to-leading-order (NLO) and consecutively added to that $1/m_b^2$ and $1/m_b^3$ terms. In addition, the crosses indicate the MC-results in the BLNP and DFN model as described in the text.}
  \label{fig::emom}
\end{figure}

\begin{figure}[h]
	\centering
	\subfloat{\includegraphics[width=0.5\textwidth]{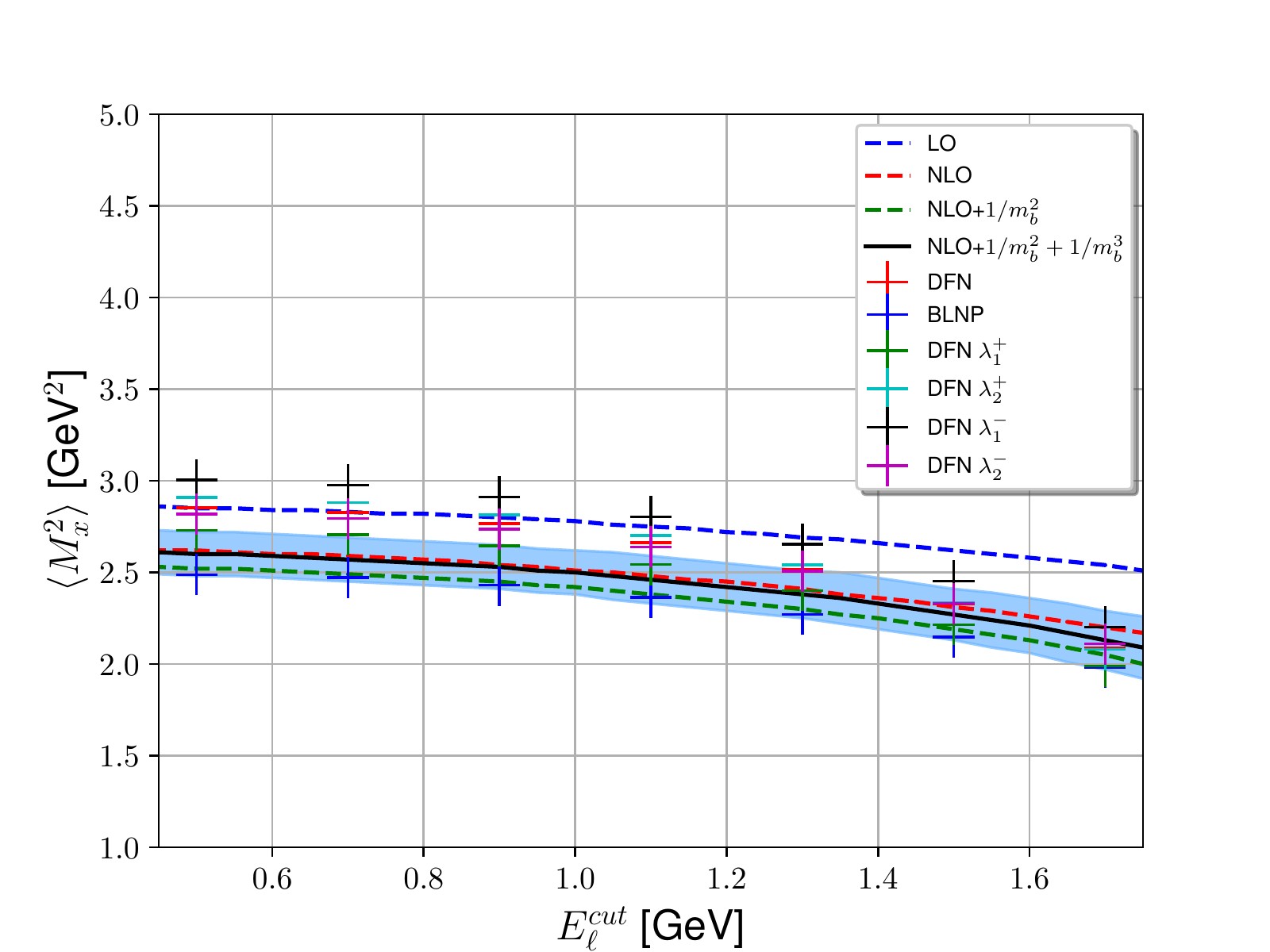}} 
  	\subfloat{\includegraphics[width=0.5\textwidth]{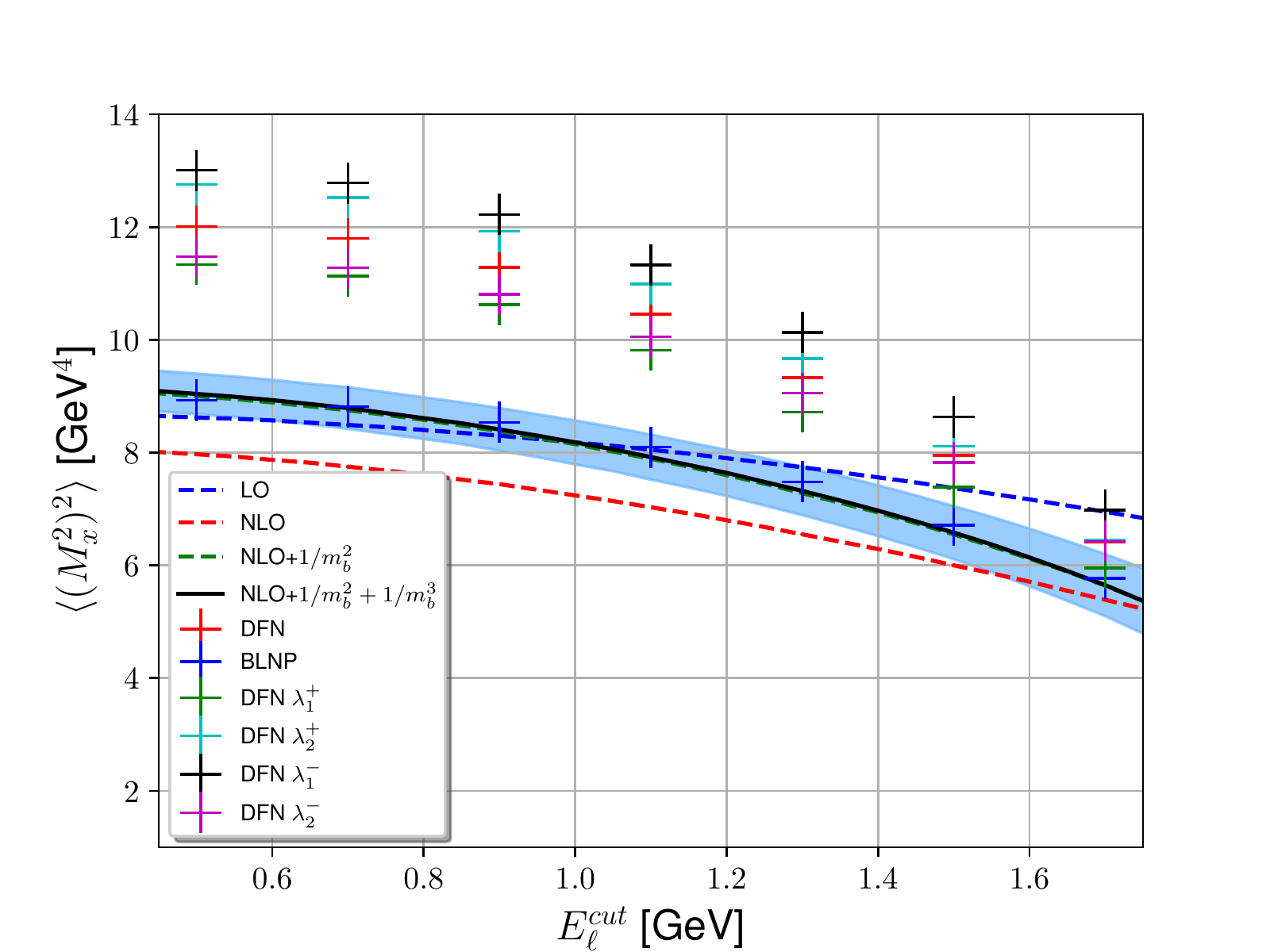}} \\
  	\subfloat{\includegraphics[width=0.5\textwidth]{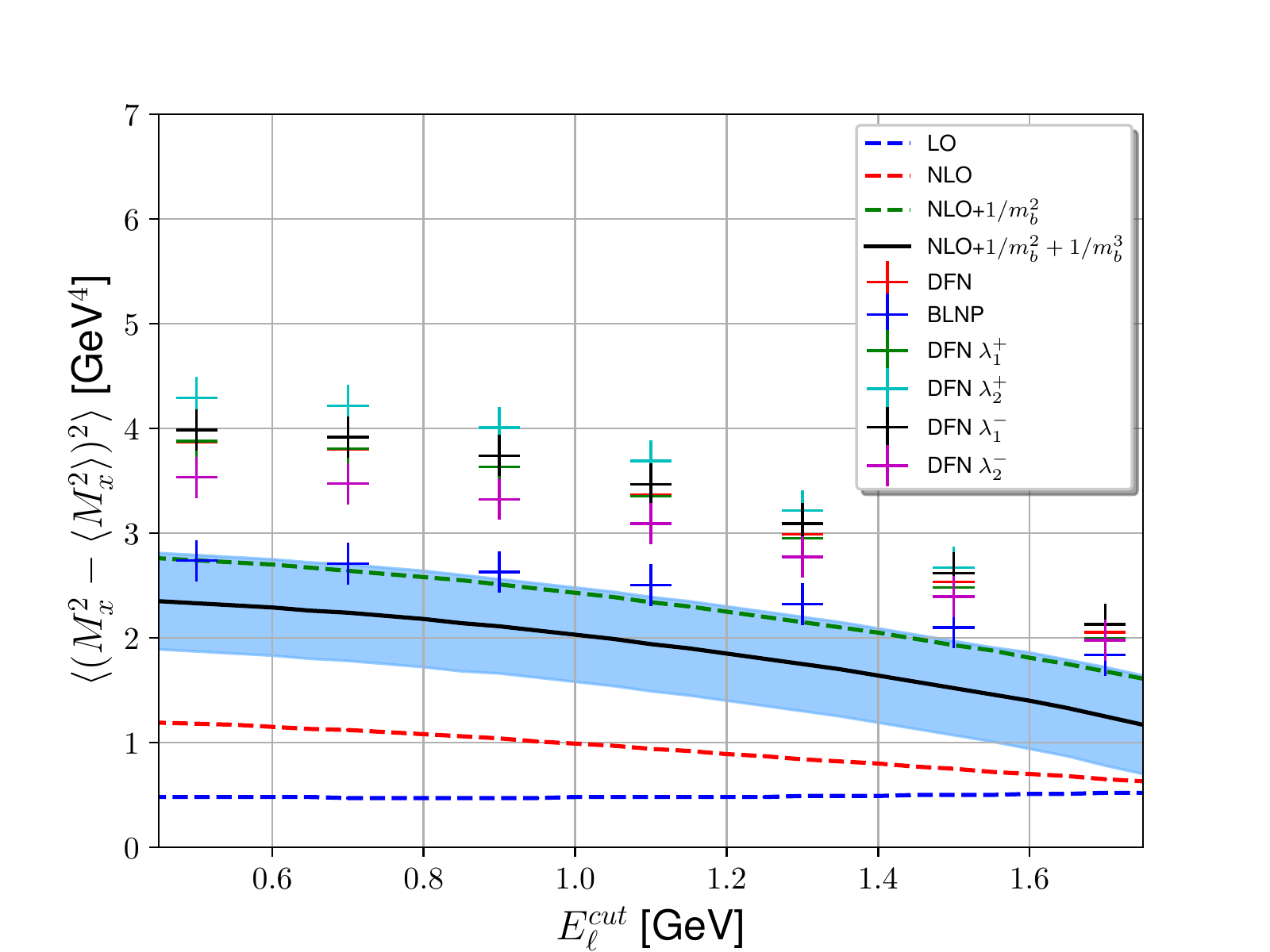}}
  	\subfloat{\includegraphics[width=0.5\textwidth]{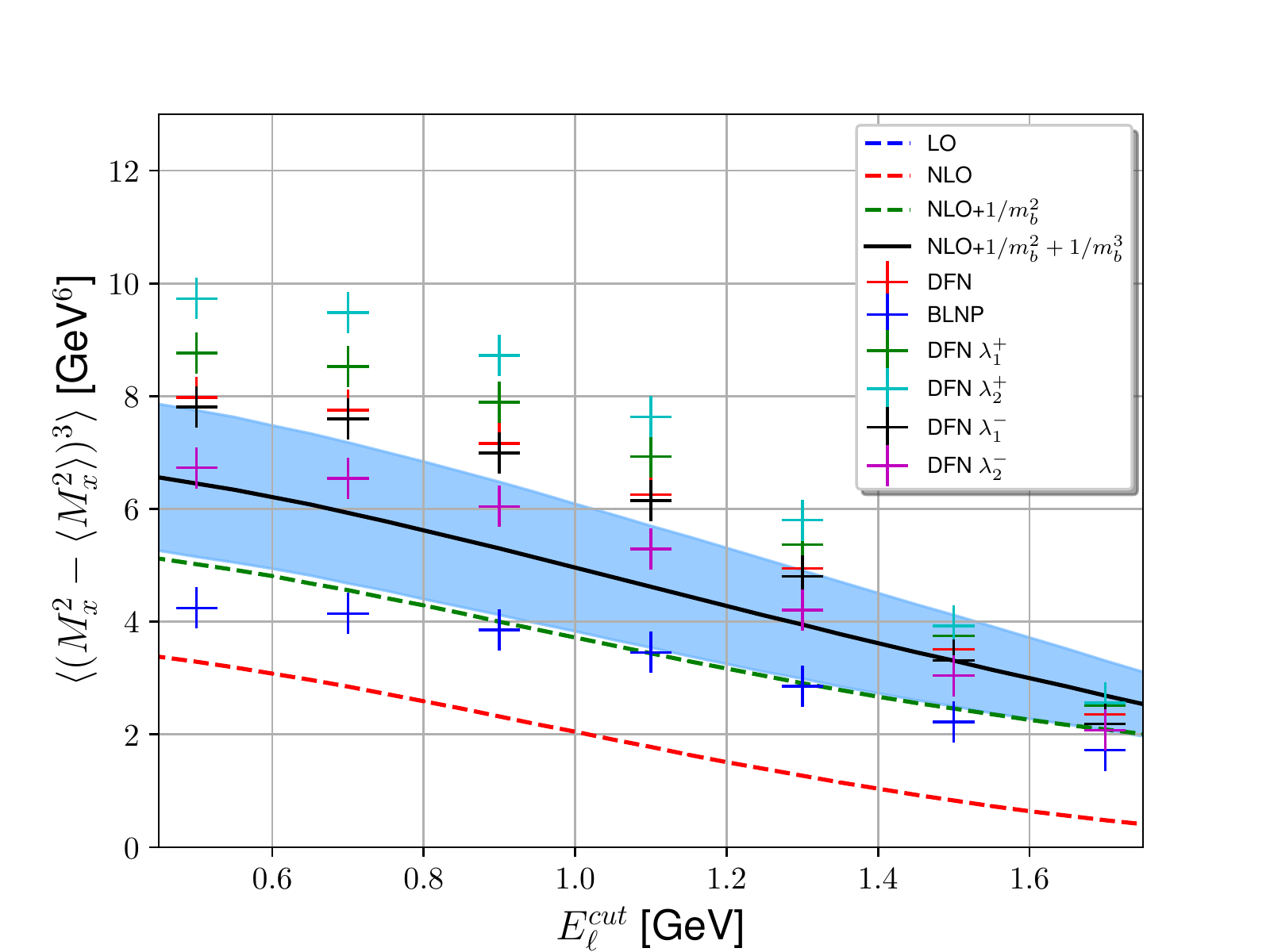}} 
  \caption{HQE results for hadronic-mass moments at different energy cuts compared with MC-results. See text and Fig.~\ref{fig::emom} for explanation.}
  \label{fig::mxmom}
\end{figure}

\begin{figure}[h]
	\centering
	\subfloat{\includegraphics[width=0.5\textwidth]{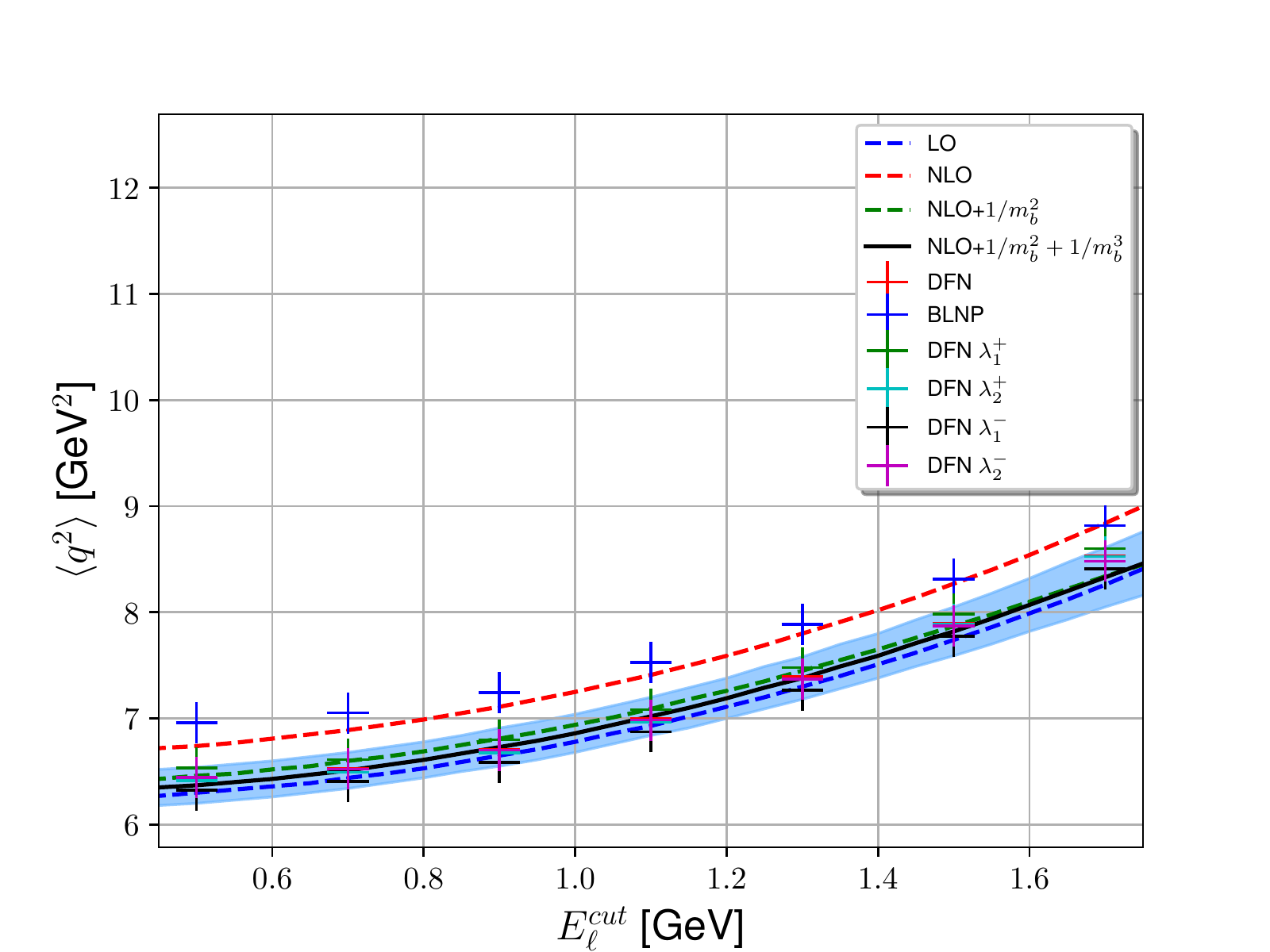}}
	\subfloat{\includegraphics[width=0.5\textwidth]{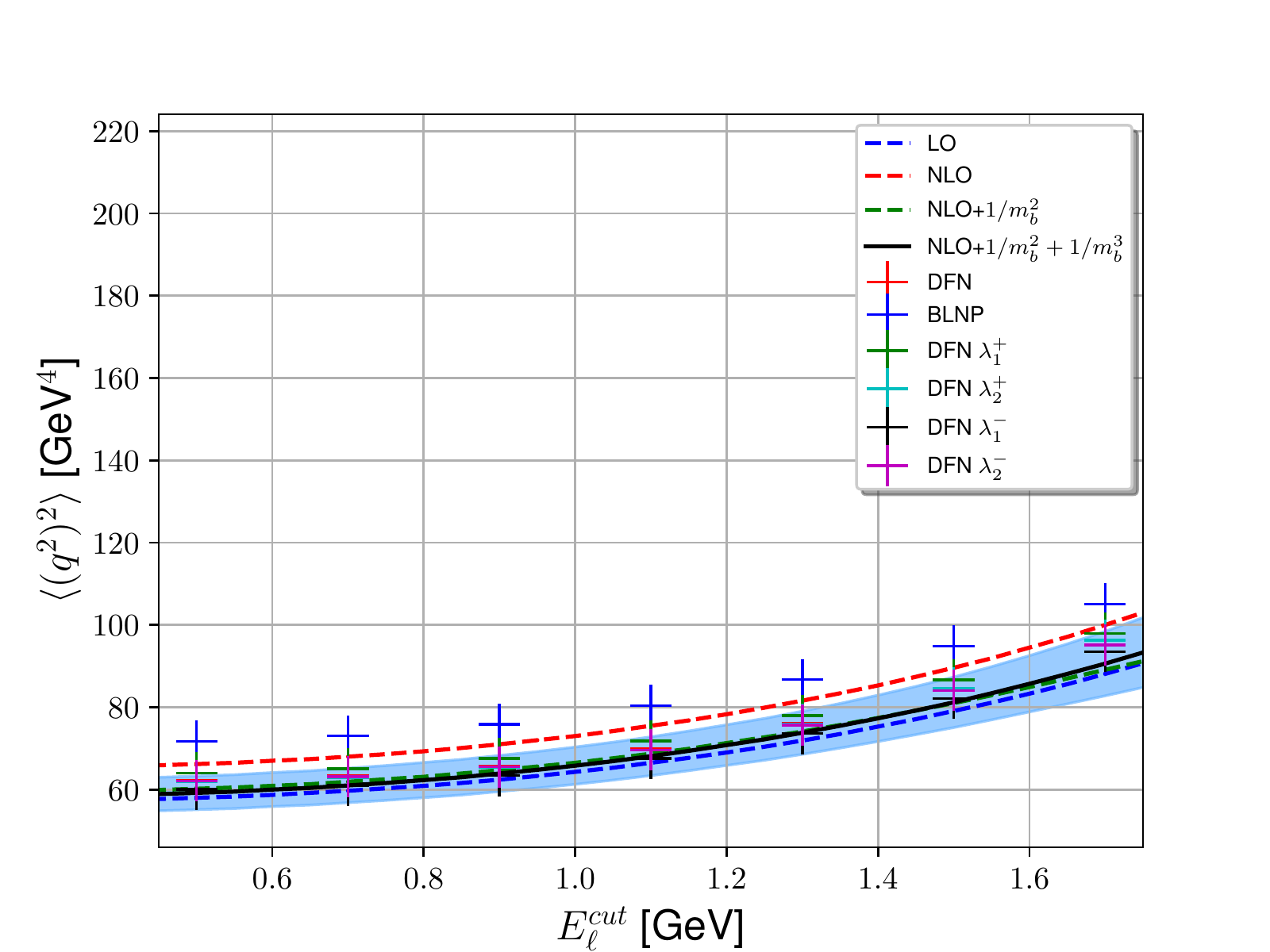}} \\
  	\subfloat{\includegraphics[width=0.5\textwidth]{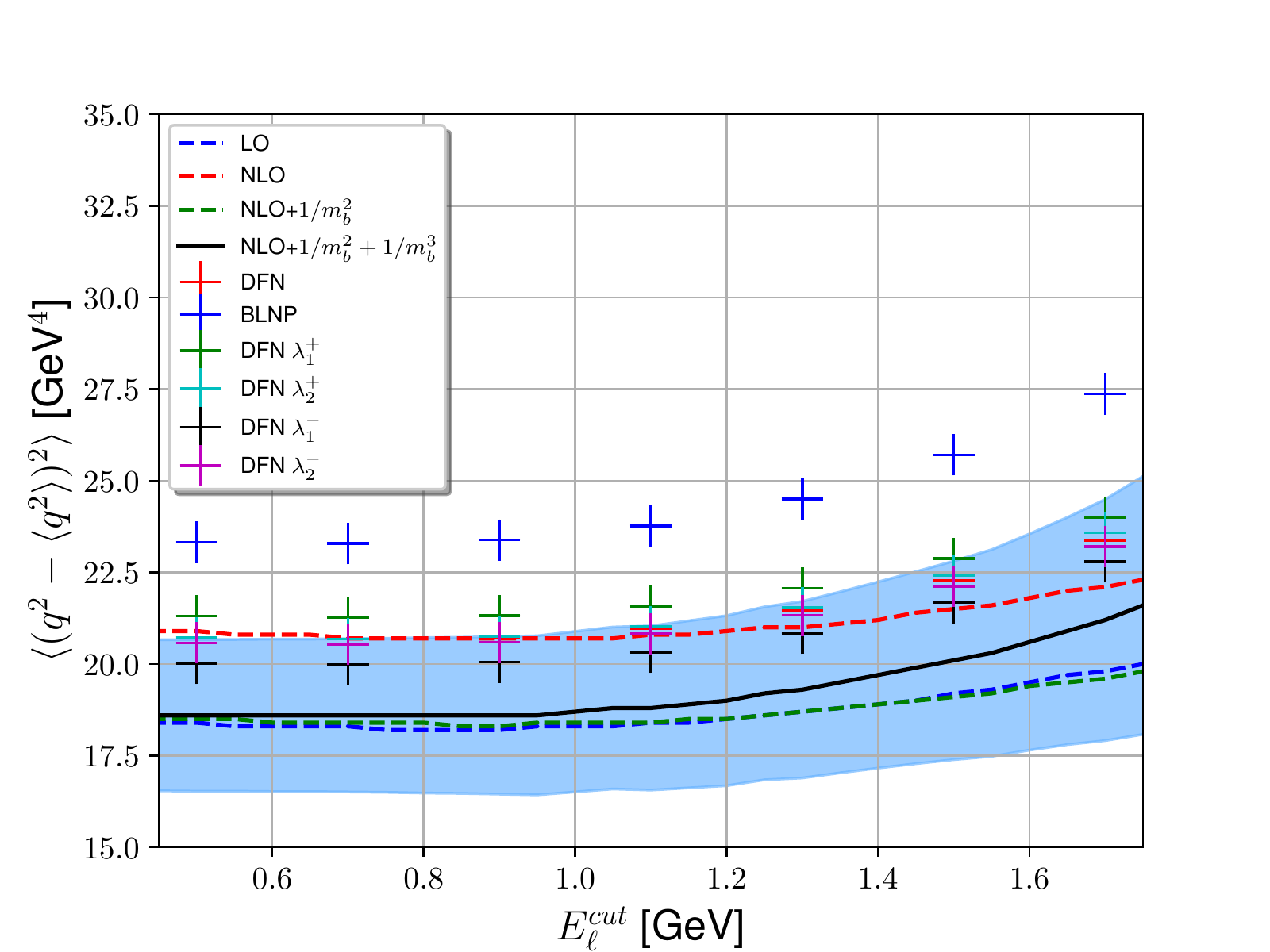}} 
  	\subfloat{\includegraphics[width=0.5\textwidth]{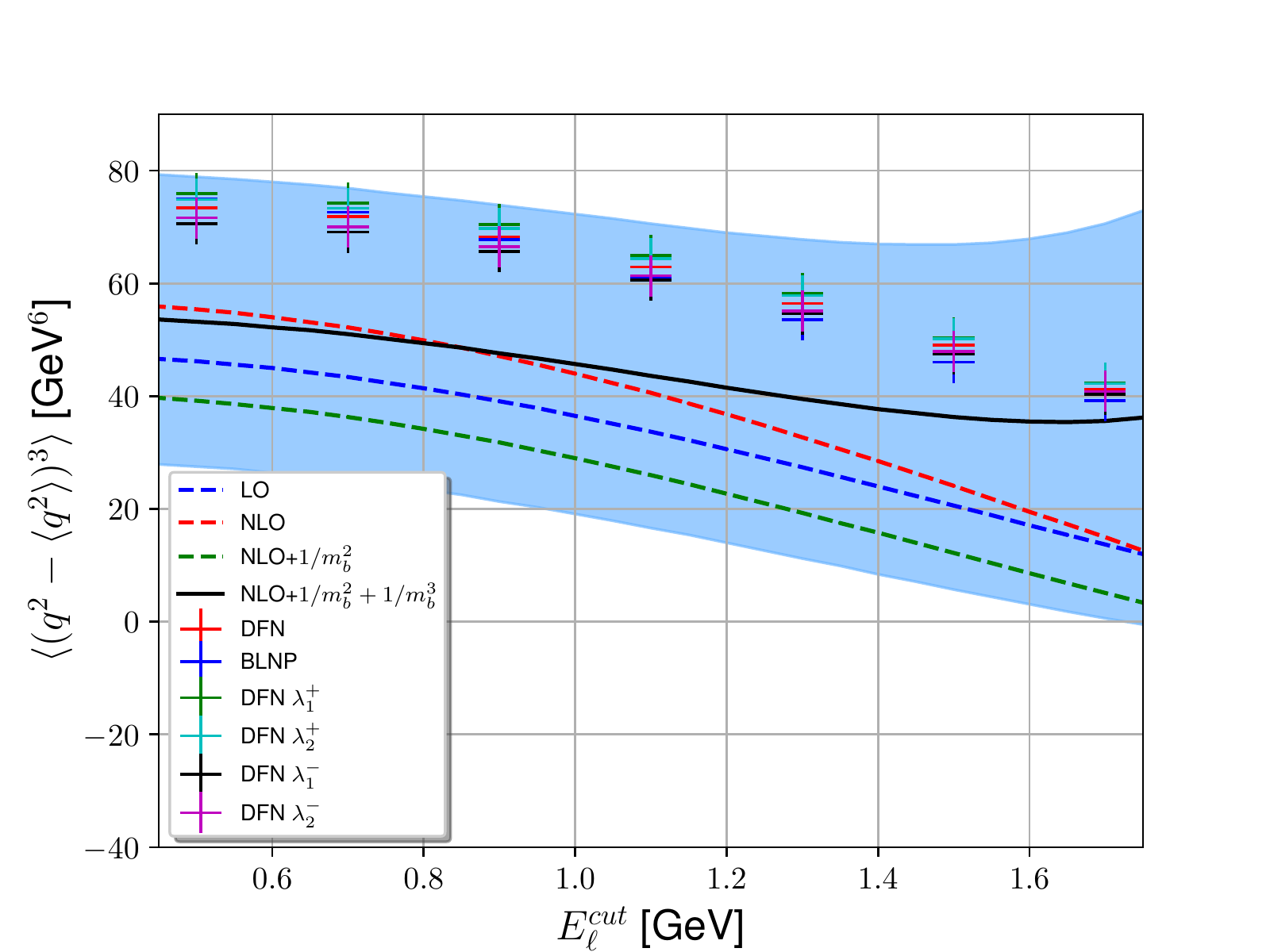}} 
  \caption{HQE results for $q^2$ moments with different energy cuts compared with MC-results. See text and Fig.~\ref{fig::emom} for explanation.}
  \label{fig::q2mom}
\end{figure}

In Figs.~\ref{fig::emom}, \ref{fig::mxmom} and \ref{fig::q2mom}, we show respectively the $E_\ell$, the $M_x$ and $q^2$-moments. For these observables, we show the first moment ($\braket{E_\ell}$), the second moment ($\braket{E_\ell^2}$) and the central moments: ($\braket{ (E_{\ell} - \braket{E_{\ell}})^{2}}$) and ($\braket{ (E_{\ell} - \braket{E_{\ell}})^{3}}$) (and equivalently for $M_x$ and $q^2$ moments). For our OPE results we show leading-order (LO), next-to-leading-order (NLO) and NLO plus $1/m_b^2$ and $1/m_b^3$ power-corrections individually. The NLO $+ 1/m_b^2 + 1/m_b^3$ is our final results, for which the blue band indicates the uncertainty obtained by varying the input parameters in Table~\ref{table::input}. We do not attribute an additional uncertainty for missing higher-order terms. These theoretical predictions are then compared to generator-level Monte-Carlo (MC) results\footnote{We thank F. Bernlochner and L. Cao for providing us these generator-level MC results which were obtained using \cite{genmc}}. 

The crosses indicate the MC data points for several methods. The MC data point labelled BLNP uses the BLNP \cite{Lange:2005yw} description of the $B\to X_u \ell \nu$ spectrum where for the input parameters of the shape function $b=3.95$ and $\Lambda=0.72$ are used.  Besides, we show 5 points labelled DFN which is based on \cite{DeFazio:1999ptt}. This DFN model contains $\alpha_s$ corrections convoluted with the non-perturbative shape function in an ad-hoc exponential model \cite{Kagan:1998ym}. The two parameters of this shape function in the Kagan-Neubert scheme are taken from a fit to $B\to X_c \ell \nu$ and $B\to X_s \gamma$ data \cite{Buchmuller:2005zv} (see also \cite{Cao:2021xqf}). In the figures, the points labelled DFN present the central values of the DFN, while $(\lambda_{1}^{+}, \lambda_{2}^{+}, \lambda_{1}^{-},\lambda_{2}^{-})$ are obtained by varying $\bar{\Lambda}$ and $\mu_\pi^2$ within $1\sigma$ regions obtained in \cite{Buchmuller:2005zv}. These variations can be used to estimate the error of the DFN model. This method, using the variation of the DFN models as an error, is used at Belle (see also the recent Belle analysis of the $\bar{B}\to X_u \ell \bar\nu$ \cite{Cao:2021xqf}). For both the DFN and the BLNP models, resonant contributions are included using a  ``hybrid Monte Carlo''. This method is based on the partonic calculation described above convoluted with a hadronization simulation based on \textsc{Pythia}, combined with $\bar{B} \to \pi \ell \bar{\nu}$ and   $\bar{B} \to \rho \ell \bar{\nu}$ at small invariant partonic invariant masses (see \cite{Prim:2019gtj,PhysRevD.41.1496}). 


Comparing our results with the MC-generated results, we observe:
\begin{itemize}
    \item {\bf for energy moments (Fig.~\ref{fig::emom}):} MC-results are in good agreement with the HQE results. However, we observe a slight deviation from the central values for the second and third central moments. It is known, that central moments are sensitive to non-perturbative effects, and thus we conclude that this small deviation indicates that the MC does not properly incorporate the non-perturbative effects.
    \item {\bf for hadronic mass moments (Fig.~\ref{fig::mxmom}):} We observe that the MC 
    results exhibit a large spread which is significantly larger than the uncertainty of the 
    HQE prediction, in particular for small lepton energy cuts. Especially the higher central 
    moments are sensitive to non-perturbative effects, which indicates that the models implemented
    in the MC do not properly describe the non-perturbative aspects.   
    \item {\bf for $q^2$ moments (Fig.~\ref{fig::q2mom}):} The DFN models agrees well with the HQE result within the estimated uncertainty. However, the BLNP model agrees well with the HQE up to $\mathcal{O}(\alpha_s)$. Similarly to the central moments of the hadronic invariant mass, the central moments of $q^2$ are deviating from the OPE result. Especially for the third central moment the BNLP model and DFN models are spreading quite far away showing again that the non-perturbative effects are not properly included. 
\end{itemize}

A few extra comments concerning the Monte Carlo models should be made. The DFN model mainly relies on perturbation theory (up to a smearing corresponding 
	to a shape function, mimicking some non-perturbative effects), and thus it is not surprising 	that these models have difficulties to capture the non-perturbative contributions that are properly described in the HQE. However, the BLNP approach can in principle properly describe the 
	results of the HQE, provided that its parameters are adjusted to the local HQE (which we use here). This requires including also the higher moments of the shape-function model as well including subleading shape functions, again with properly adjusted moments. The visible deviation of BLNP from the HQE predictions indicates that the version of BLNP employed in 
	the MC should be updated.

\clearpage
\section{Background from the $\bar B\to X_c  (\tau \to \ell \bar\nu_\ell \nu_\tau) \bar\nu_\tau$ decay}
\label{ch::Inclusive-five-body}

Next we consider the background contribution of $b \to c \tau (\to \ell \bar{\nu}_\ell \nu_\tau) \bar{\nu}_{\tau}$. Similar to the $b\to u$ background, this five-body contribution can be calculated exactly within the HQE. Its contribution is expected to be small, due to phase space suppression and the small branching fraction of $\tau \to \ell  \bar{\nu}\nu$. Experimentally this five-body contribution can be further reduced by for example cutting on the lepton momentum in the $B$ restframe and by constraining the invariant mass of the $B$-meson. Due to these extra cuts, it is not as straightforward to compare our exact HQE results with the experimentally used Monte-Carlo data as for the $b\to u$ contamination. In the following, we present our theoretical calculations of this five-body contribution, which may be used to improve the description of this background. 

For our HQE calculation, we proceed by following \cite{Bordone:2016tex} where the $\tau$-contribution to exclusive $B\to D \ell \nu$ decay was studied (see also \cite{Alonso:2016gym}). To our knowledge, the $\tau$ contribution to inclusive decays was never studied. Semitauonic $B$ decays were studied in \cite{Mannel:2017jfk}.

In order to obtain lepton energy, hadronic mass and $q^2$ moments, we construct the differential decay rate of the $B(p_B)\to X_c (p_{X_c}) (\tau (q_{[\tau]})\to \mu (q_{[\mu]}) \nu_\mu (q_{[\bar{\nu}_\mu]}) \nu_\tau (q_{[\nu_\tau]})) \bar{\nu}_\tau (q_{[\bar{\nu}_\tau]})$:
\begin{align}
  &  \frac{\text{d}^8 \Gamma}{\text{d} q^2 \, \text{d} q^2_{\nunubox} \, \text{d}p^2_{X_{c}} \, \text{d}^2 \Omega \, \text{d} \Omega^{*} \, \text{d}^2 \Omega^{**}} = \nonumber \\
    &- \frac{3 G_F^2 |V_{cb}|^2 \sqrt{\lambda} (q^2 - \mtau^2)(\mtau^2 - q^2_{\nunubox}) \mathcal{B}(\tau \to \mu \nu \nu)}{2^{17} \pi^5 \mtau^8 m_b^3 q^2} W_{\mu \nu} L^{\mu \nu} \, ,
    \label{eq::fully-differential-decay-width}
\end{align}
with $q^2_{\nunubox} = (q_{\barnumubox} + q_{\nutaubox})^2$, $\text{d}^2 \Omega = \text{d} \cos{ \theta_{\taubox}} \text{d} \phi$, $\text{d} \Omega^* = \text{d} \cos{\theta^{*}_{\mubox}}$, $\text{d}^2 \Omega^{**} = \text{d} \cos{\theta^{**}_{\barnumubox}} \text{d} \phi^{**}$ and $\lambda \equiv \lambda(m_b^2, m_c^2, q^2)$ is the K\"all\'en-function. For the different angles we follow the conventions in \cite{Bordone:2016tex}. 
The lepton tensor is now given by
\begin{equation}
    L^{\mu\nu} = \sum_{\rm spins} L^\mu L^{\nu *} \ ,
\end{equation}
with
\begin{align}
    L^{\mu} &= \frac{1}{q_{\taubox}^2 -\mtau^2 + i \mtau \Gamma_\tau} \left [ \bar{u}(q_{\mubox}) \gamma_\alpha (1-\gamma_5) v(q_{\barnumubox}) \right] \nonumber \\
    & \times \left[ \bar{u}(q_{\nutaubox}) \gamma^{\alpha} (1- \gamma_5)(\slashed{q}_{\taubox} + \mtau) \gamma^\mu (1-\gamma_5) v(q_{\barnutaubox}) \right] \, .
\end{align}
For the $\tau$, we use the narrow-width approximation $\Gamma_\tau \ll \mtau$:
\begin{align}
   \left| \frac{1}{(q^2_{\taubox} - \mtau^2 + i \mtau \Gamma_{\tau})}\right|^2 &\underset{\Gamma_\tau \ll \mtau}{\longrightarrow}  \frac{\pi}{\mtau \Gamma_\tau} \delta(q^2_{\taubox} - \mtau^2 ) \ ,
\end{align}
where $\Gamma_\tau$ is the total width of the $\tau$ lepton.
We want to obtain moments of the differential spectrum with cuts on the lepton energy as before. The lepton energy of the muon $E_\ell$ can be related to $q^2_{\nunubox}$ in the following way:
\begin{align}
    E_\ell &= \frac{1}{2 m_{b}} \beta _{\nu \bar\nu } \left(\sqrt{\lambda } \sqrt{1-2 \beta _{\tau }} \sin
   \theta_{\taubox} \sin \theta^{*}_{\mubox} \cos\phi -\sqrt{\lambda } \cos \theta_{\taubox} \nonumber \right. \\ 
   & \left. \times \left(\left(1-\beta _{\tau }\right) \cos\theta^{*}_{\mubox} +\beta _{\tau}\right)+\left(m_b^2-p_{X_{c}}^2 + q^2\right) \left(\beta _{\tau } \cos\theta^{*}_{\mubox}-\beta _{\tau }+1\right)\right)
\end{align}
with
\begin{align}
    \beta_{\nu\bar\nu} &= \frac{\mtau^2 -q^2_{\nunubox}}{2 \mtau^2} \,  \hspace{1cm} \text{and} \hspace{1cm} \beta_{\tau} = \frac{q^2 -\mtau^2}{2 q^2} \, .
\end{align}

The five-body phase-space is similar to the exclusive decay, but the contraction of the hadronic tensor $W_{\mu\nu}$ with the leptonic tensor $L_{\mu\nu}$ differs from the exclusive decay. We discuss the leptonic tensor and definitions of the four-vectors in more detail in our upcoming work~\cite{WorkInProgress3}.  The hadronic tensor $W_{\mu \nu}$ can be constructed following the procedure given in \cite{Dassinger:2006md} (see Eq.~\eqref{eq:tripdif}, where now also $W_{4,5}$ are relevant). Finally, we obtain the eight-fold differential decay rate. We explicitly verified that our differential rate reduces to the total decay rate of $b \to c \tau \bar\nu$, i.e.:
\begin{align}
    \Gamma_{\text{tot}}(b \to c \tau (\to \ell \bar{\nu} \nu) \bar{\nu} ) &= \Gamma_{\text{tot}}(b \to c \tau \bar\nu)\mathcal{B}(\tau \to \ell \bar\nu \nu) \, .
\end{align}
We note that the branching ratios of $\tau \to \mu\bar{\nu}\nu$ and $\tau \to e\bar\nu\nu$ are almost identical. Now we can compute the moments similarly to Sec.~\ref{ch::Inclusive-three-body} by integrating the eight differential rate over the appropriate kinematical variables. 
We do not include radiative corrections for this process. 

\begin{figure}[h]
    \centering
	\includegraphics[width=0.5\textwidth]{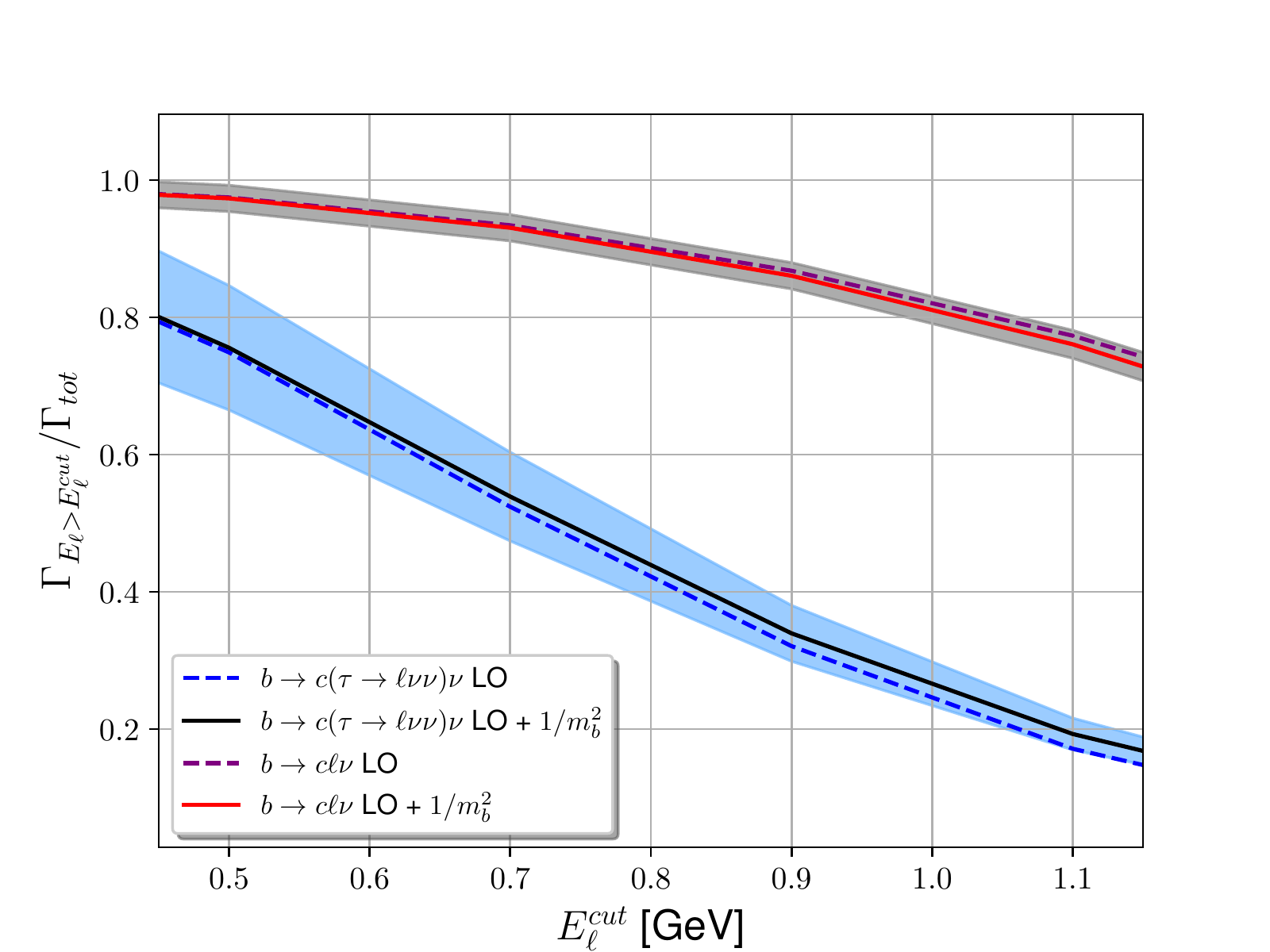}
    \caption{Five and three-body decay rates with lepton energy cut ($\Gamma_{E_{\ell} > E^{\rm cut}_{\ell}}$) normalized to the corresponding rate without energy cut ($\Gamma_{\text{tot}}$). The dotted lines indicate the LO contribution and the solid line includes $1/m_b^2$ corrections for $b\to c (\tau \to \ell \bar\nu \nu)\bar\nu$ (blue) and $b\to c \ell \bar\nu$ (red) and their uncertainty.}
    \label{fig::ZeroMom}
\end{figure}

As we mentioned before, the decay $b \to c \tau (\to \ell \bar{\nu}_\ell \nu_\tau) \bar{\nu}_{\tau}$ is small compared to $b \to c \ell \bar\nu$. Hence, the total decay rate is given as:
\begin{align}
    \frac{\Gamma_{\text{tot}}(b \to c \tau (\to \ell \bar{\nu}_\ell \nu_\tau) \bar{\nu}_{\tau})}{\Gamma_{\text{tot}}(b \to c \ell \bar\nu)} \sim 4.0 \%
\end{align}
In Fig.~\ref{fig::ZeroMom}, we show for both the five and three-body decay, the total rate with a muon energy cut $E_\ell^{\rm cut}$ normalized by the corresponding decay rate without cut, i.e. for the five-body decay:
\begin{align}
    \frac{\Gamma_{\text{tot}}(b \to c \tau (\to \ell \bar{\nu}_\ell \nu_\tau) \bar{\nu}_{\tau})|_{E_\ell>E^{\rm cut}_\ell}}{\Gamma_{\text{tot}}(b \to c \tau (\to \ell \bar{\nu}_\ell \nu_\tau) \bar{\nu}_{\tau})} \ .
\end{align}
We find that the five-body decay decreases more rapidly when increasing the lepton energy cut, which is expected as such a lepton-energy cut further reduces the already suppressed phase space. 

\begin{figure}[t]
	   \centering
        \subfloat{\includegraphics[width=0.35\textwidth]{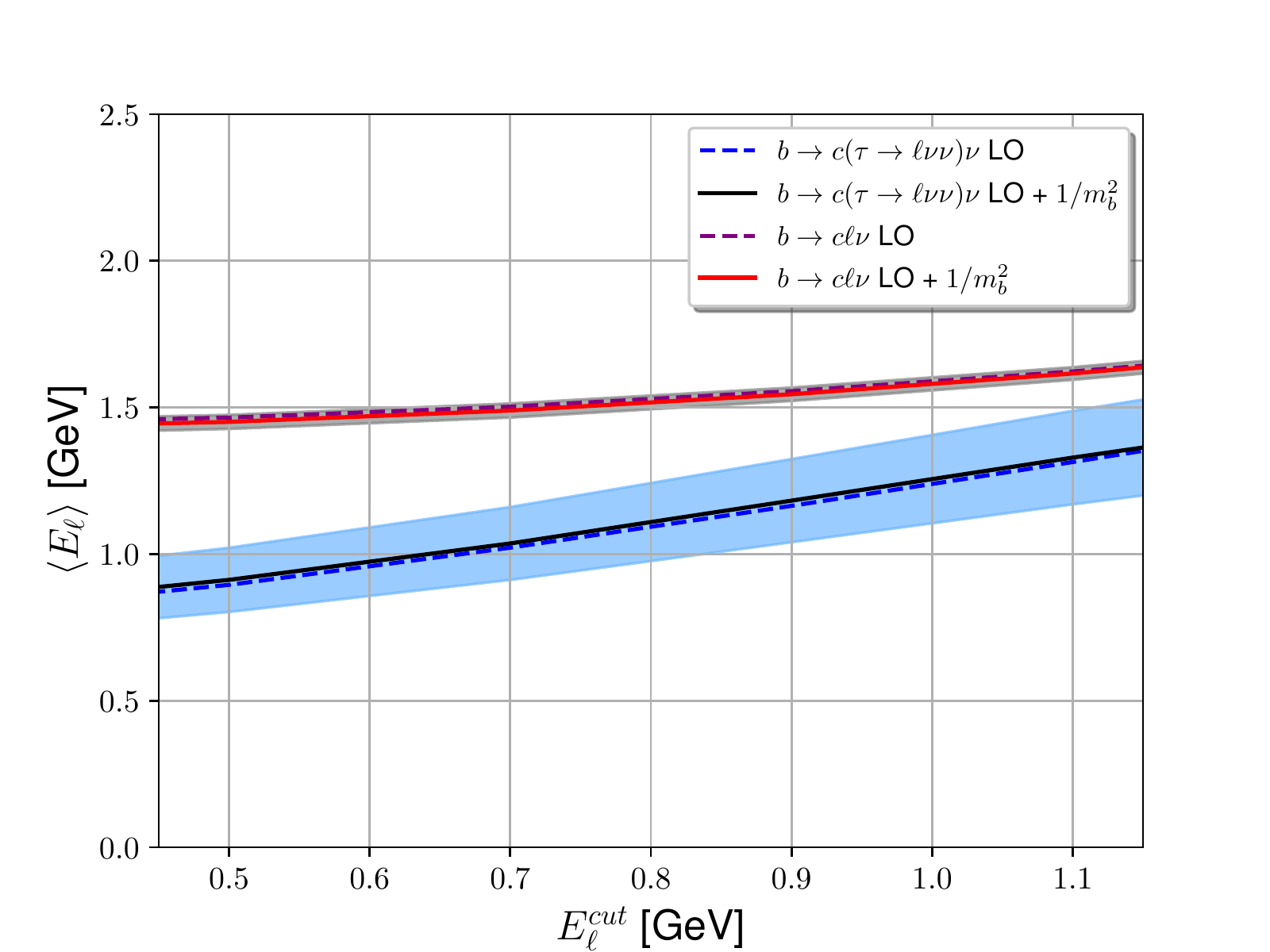}}
        \subfloat{\includegraphics[width=0.35\textwidth]{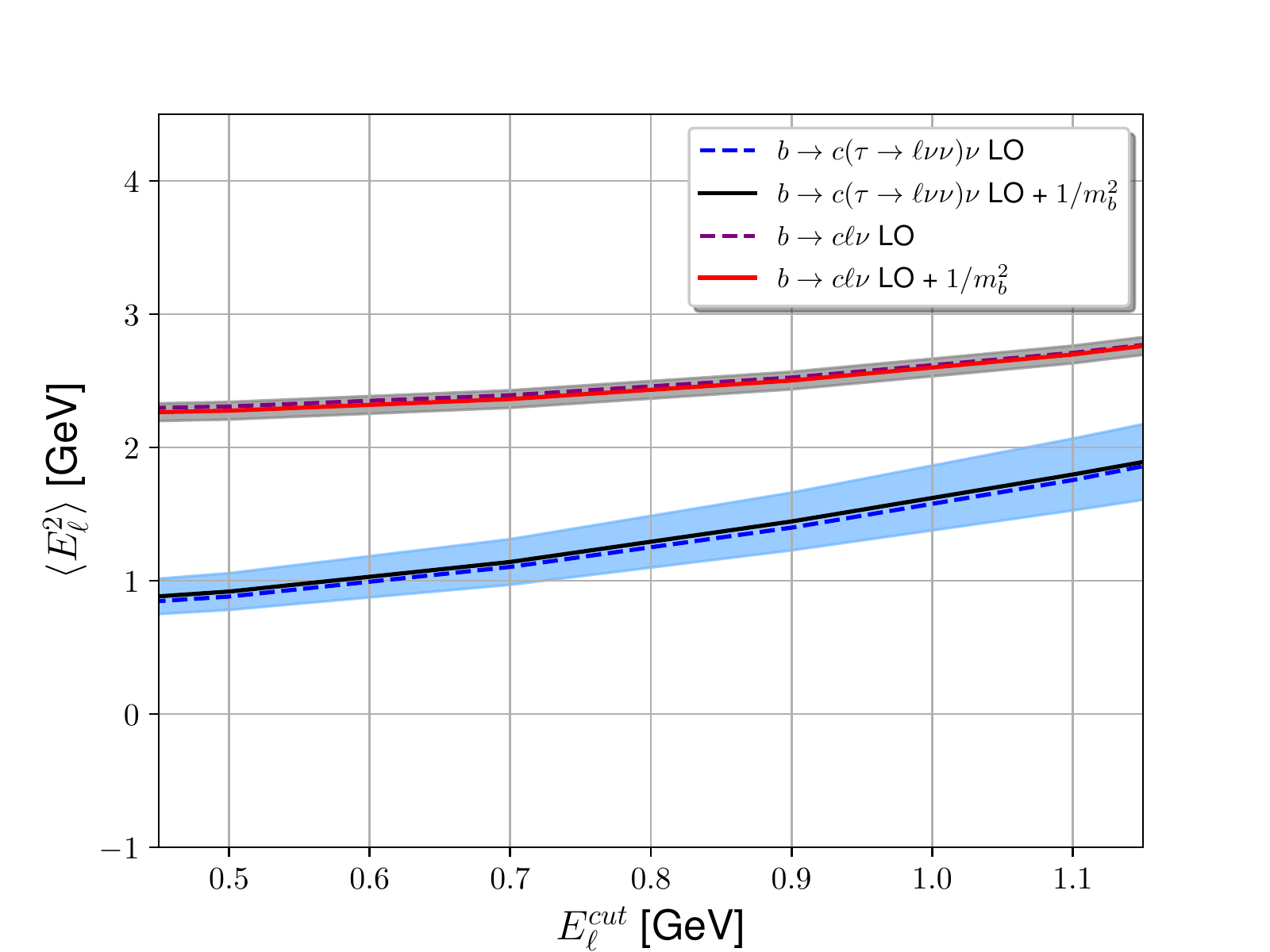}} 
        \subfloat{\includegraphics[width=0.35\textwidth]{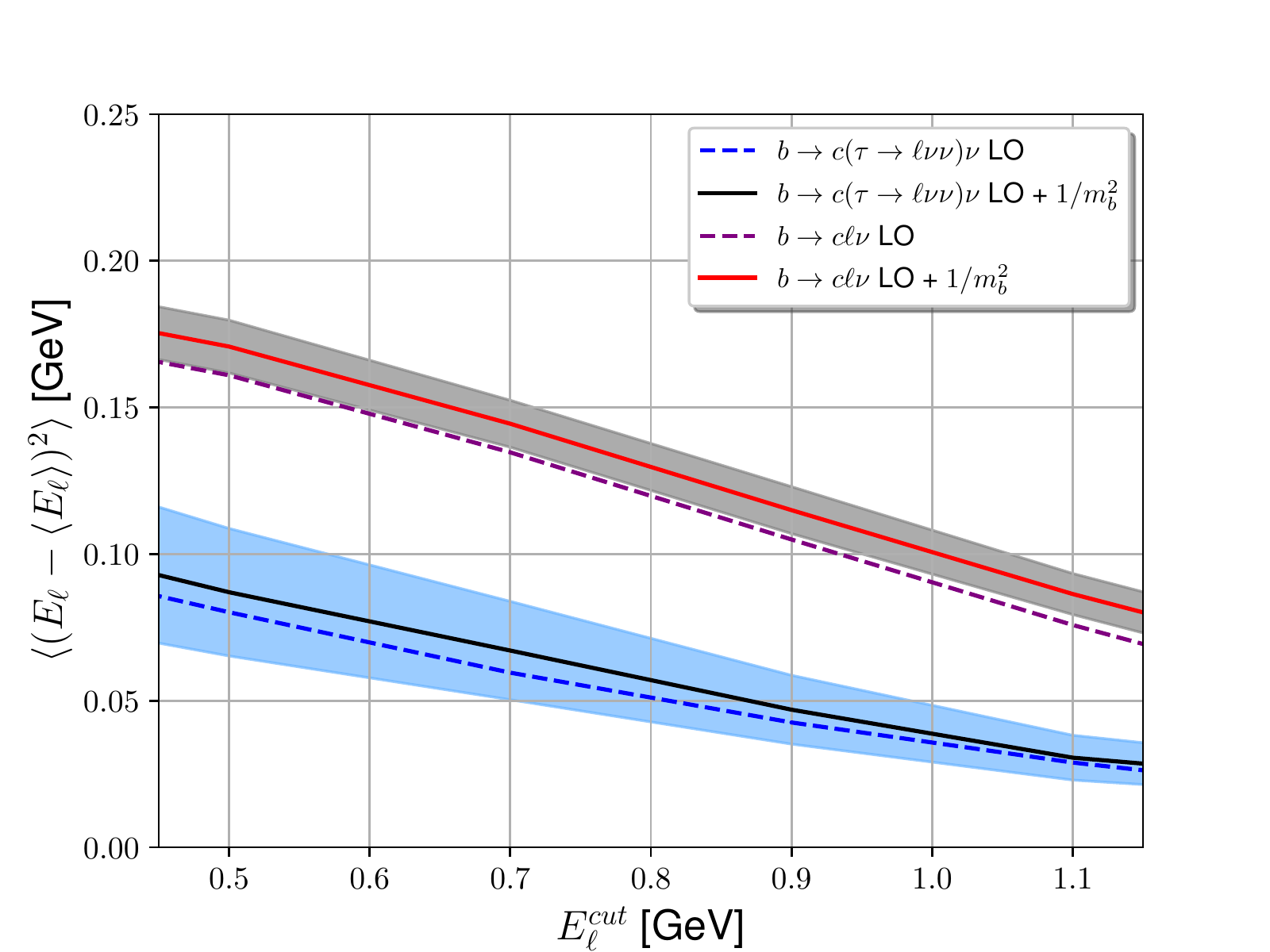}}
        \caption{$E_\ell$ moments as a function of $E_\ell^{\text{cut}}$ at LO (dotted) and including $1/m_b^2$ corrections (solid) for $b\to c (\tau \to \ell \bar\nu \nu)\bar\nu$ (blue) and $b\to c \ell \bar\nu$ (red) and their uncertainty.}
        \label{fig::EMoment_fivebody}
\end{figure}

\begin{figure}[h]
	   \centering
        \subfloat{\includegraphics[width=0.36 \textwidth]{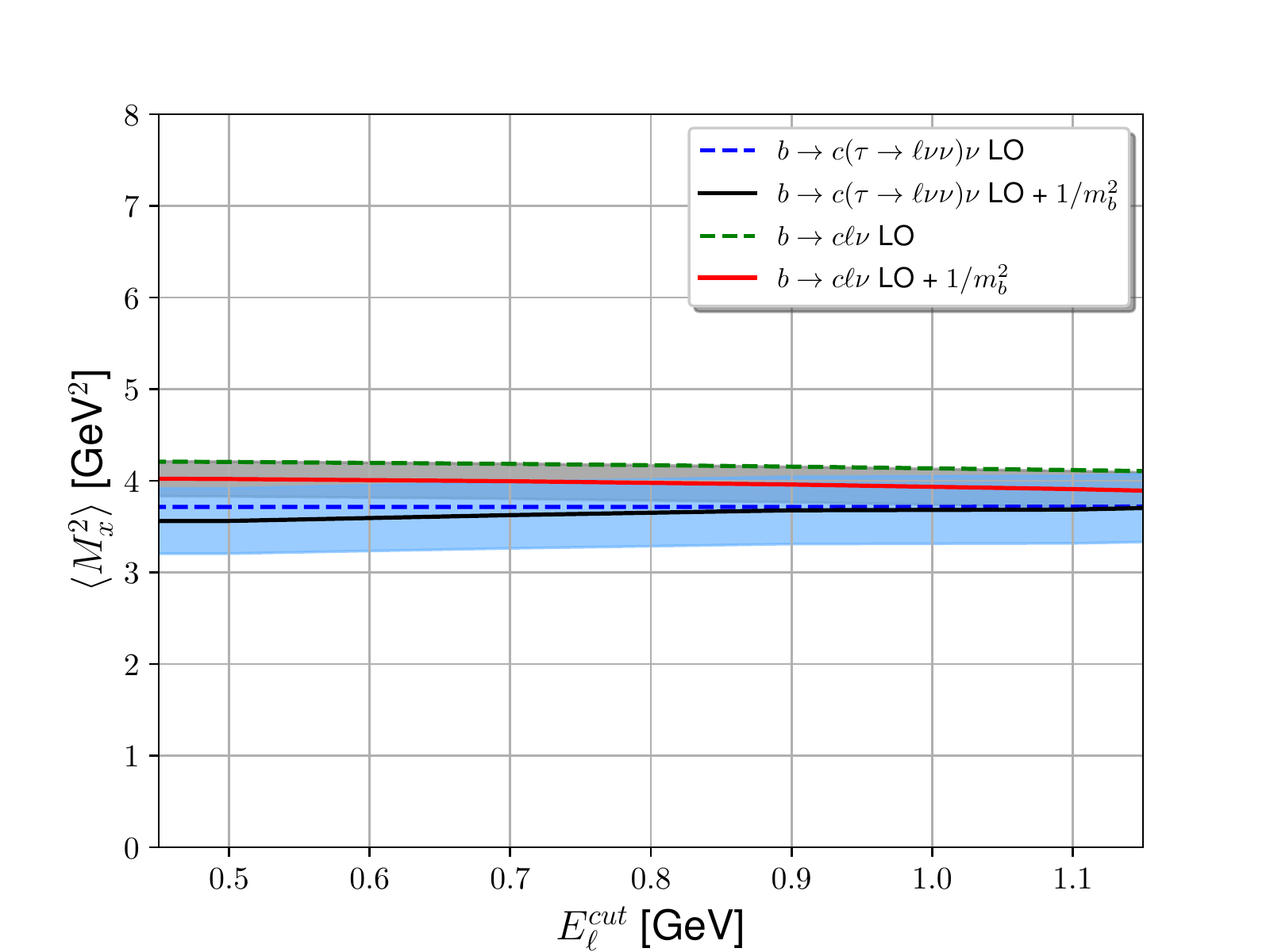}}
        \subfloat{\includegraphics[width=0.36\textwidth]{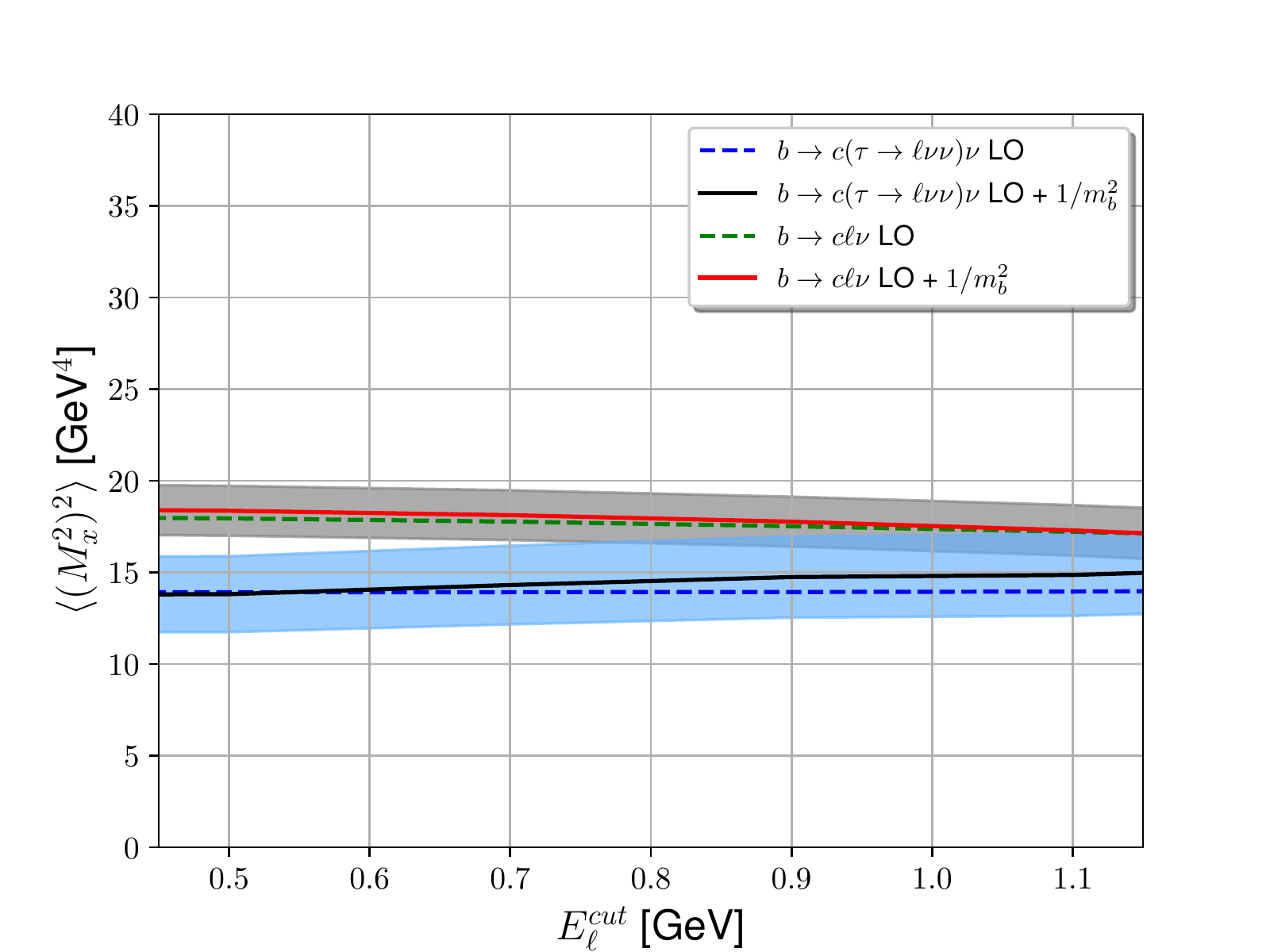}}
        \subfloat{\includegraphics[width=0.36\textwidth]{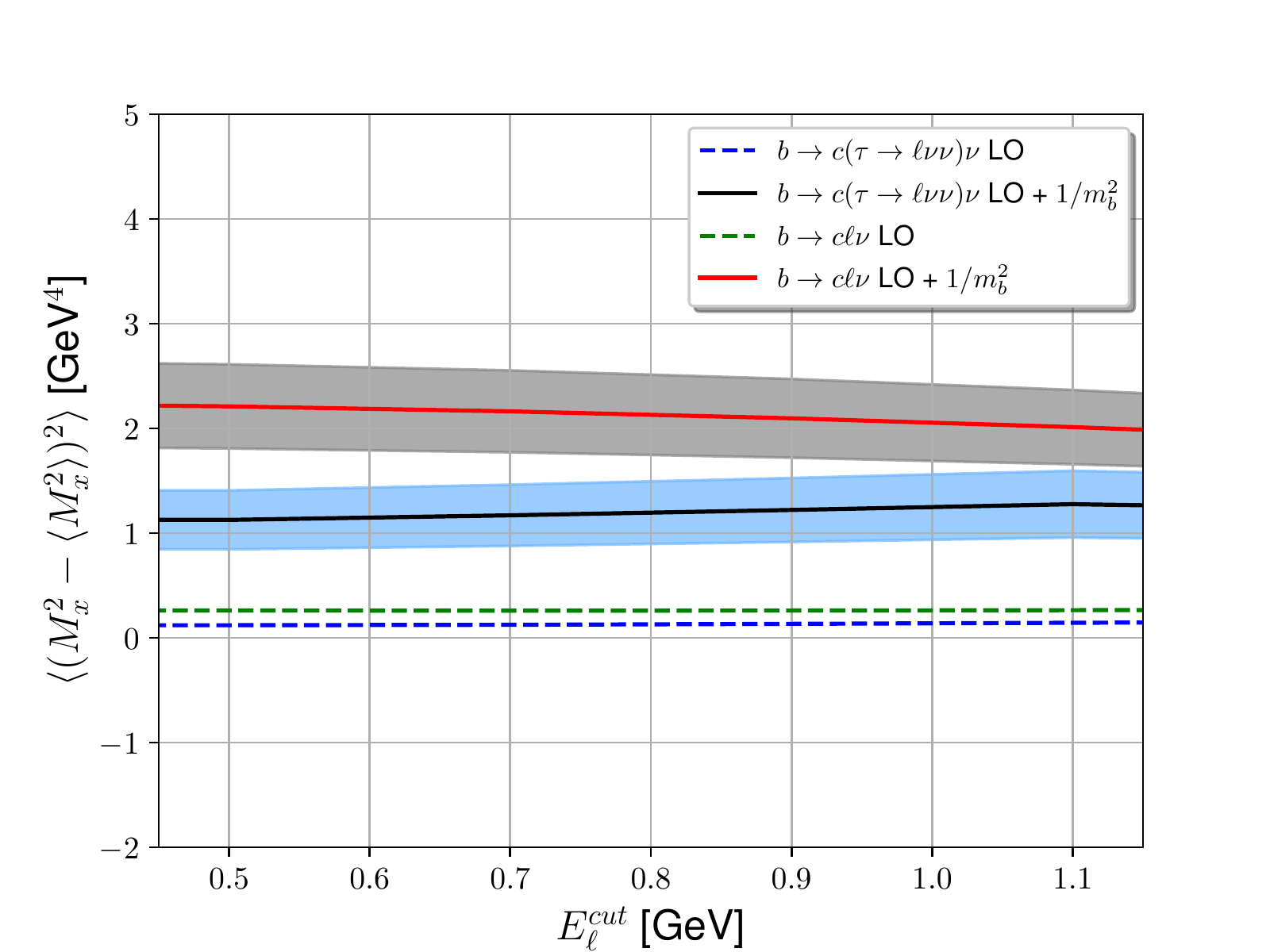}}
      \caption{$M_x^2$ moments as a function of $E_\ell^{\text{cut}}$ at LO (dotted) and including $1/m_b^2$ corrections (solid) for $b\to c (\tau \to \ell \bar\nu \nu)\bar\nu$ (blue) and $b\to c \ell \bar\nu$ (red) and their uncertainty.}
        \label{fig::MxMoment_fivebody}
\end{figure}

\begin{figure}[h]
	   \centering
        \subfloat[]{\includegraphics[width=0.36\textwidth]{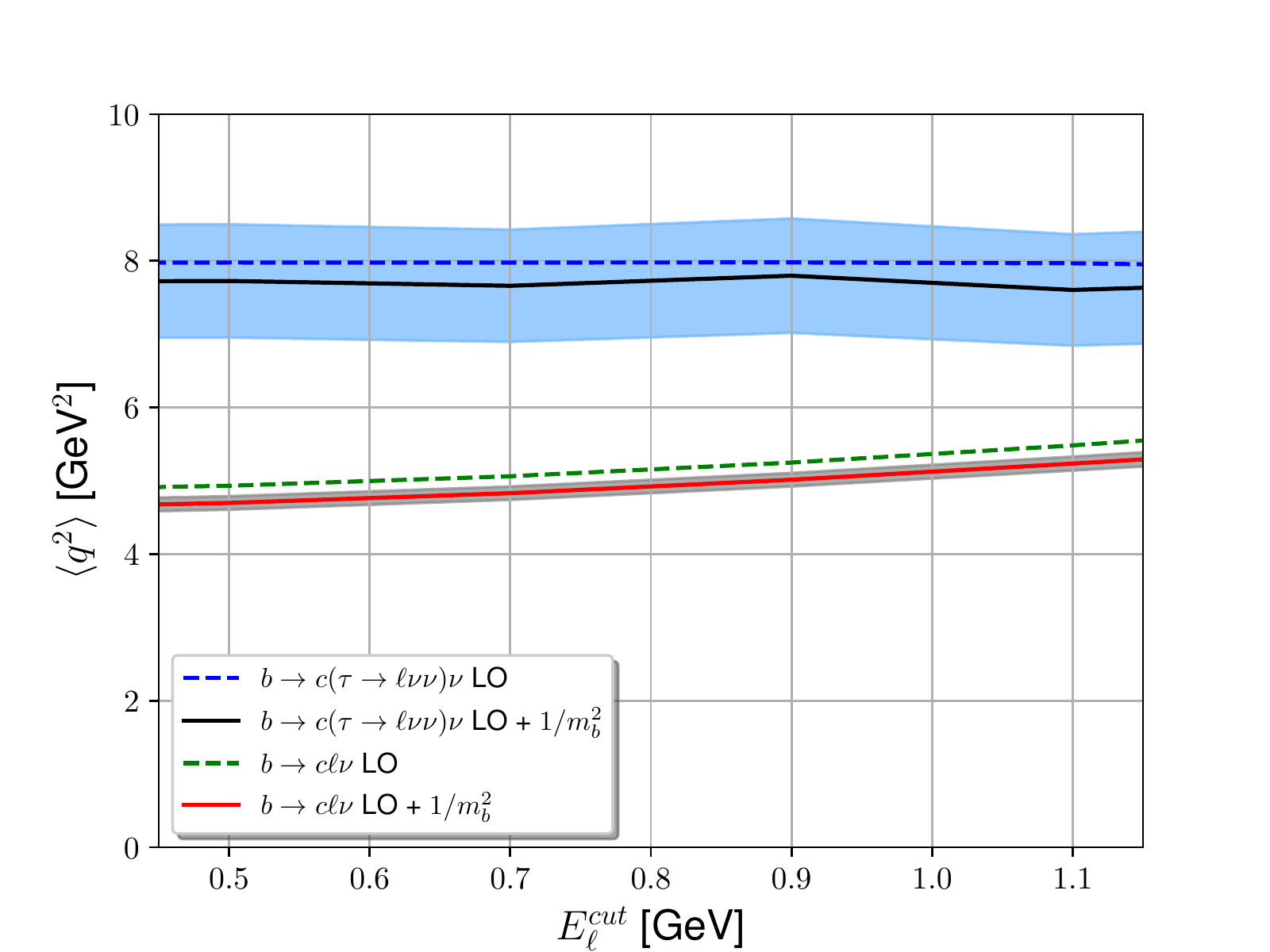}}
        \subfloat[]{\includegraphics[width=0.36\textwidth]{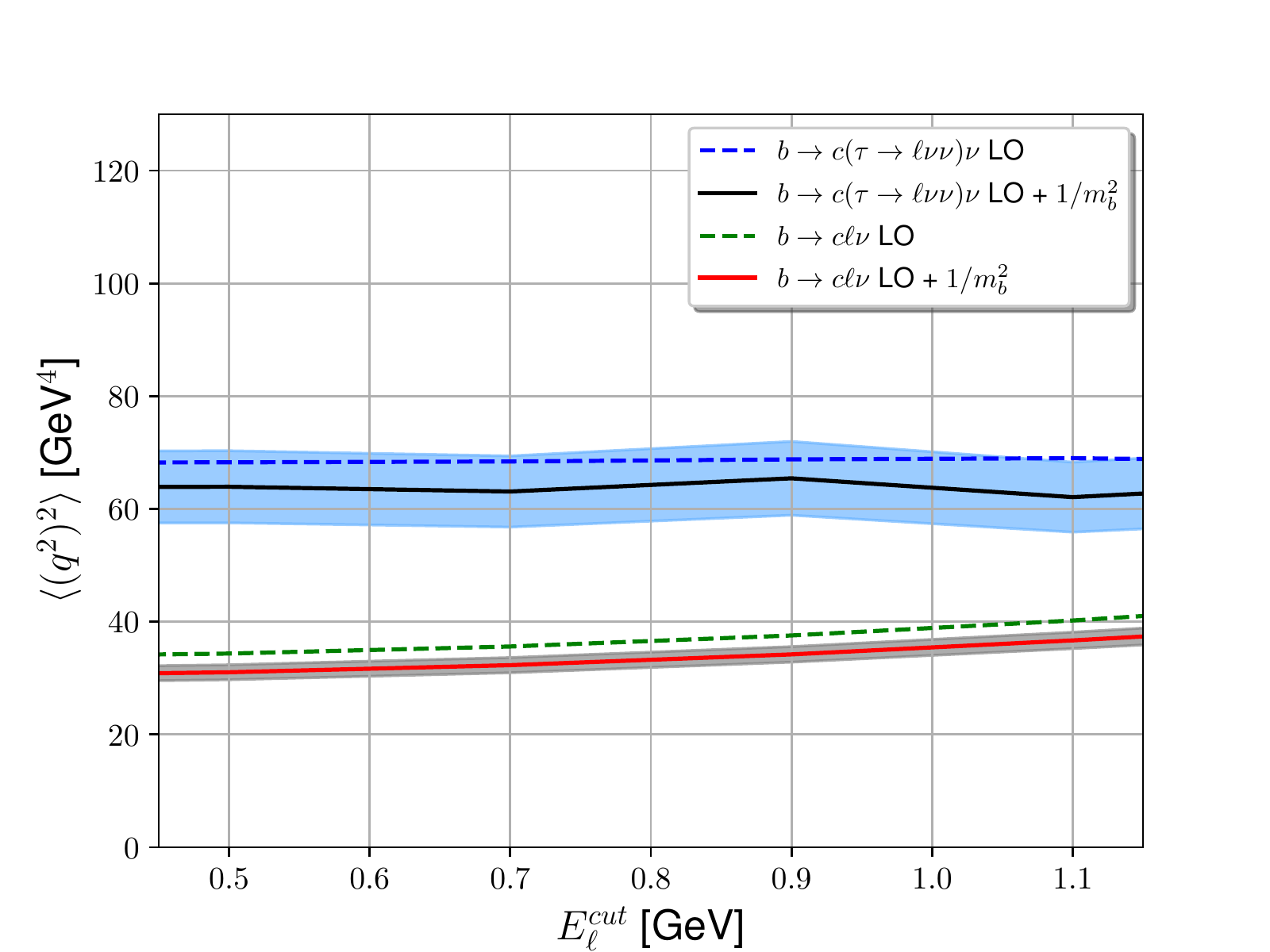}} 
        \subfloat[]{\includegraphics[width=0.36\textwidth]{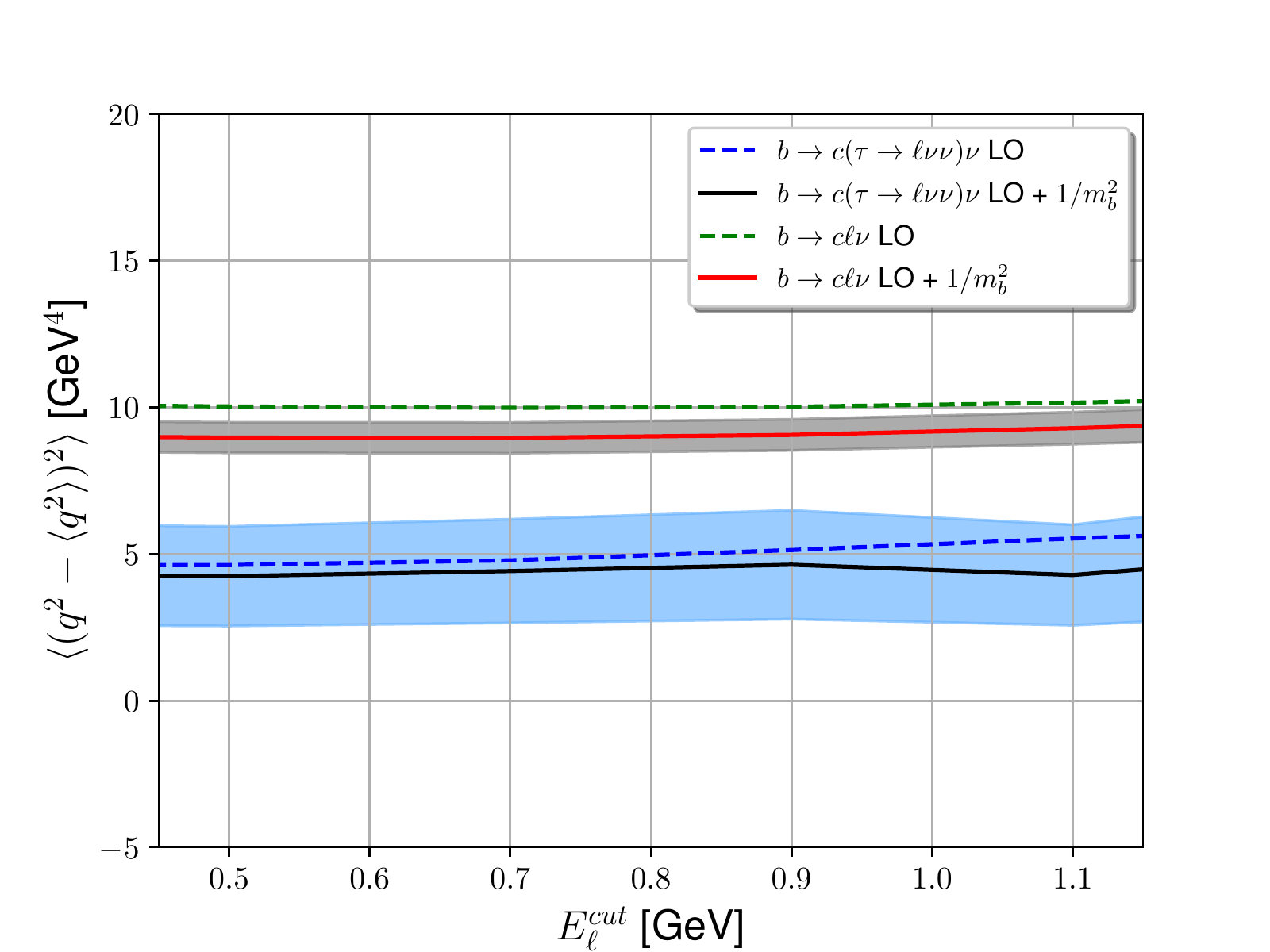}}
  \caption{The $q^2$ moments as a function of $E_\ell^{\text{cut}}$ at LO (dotted) and including $1/m_b^2$ corrections (solid) for $b\to c (\tau \to \ell \bar\nu \nu)\bar\nu$ (blue) and $b\to c \ell \bar\nu$ (red) and their uncertainty.}
        \label{fig::q2Moment1_fivebody}
\end{figure}

Finally, we obtain the lepton energy moments (Fig.~\ref{fig::EMoment_fivebody}), the hadronic mass moments (Fig.~\ref{fig::MxMoment_fivebody}) and the $q^2$ moments (Fig.~\ref{fig::q2Moment1_fivebody}). We note that we have normalized these moments to the corresponding five-body $b\to c (\tau \to \ell \bar\nu\nu)\bar\nu)$ rate. We show both the leading-order (LO, dotted line) and leading-order plus power-corrections of order $\mathcal{O}(1/m_b^2)$ (solid line). In addition, we also plot the lepton energy moments of the inclusive decay $b \to c \ell \nu$ for comparison (red). The band presents an estimate of the uncertainty obtained by varying each input parameter individually and adding them in quadrature. To compensate for missed higher-order radiative and power-corrections, we added an additional $30 \%$ uncertainty. We present these five-body moments in this way, such that they can be compared to Monte-Carlo simulations when this becomes available. 

Note that for the five-body decay $q^2 \equiv (p_\tau + p_\nu)^2 = (p_B - p_{X_c})^2$ which obviously is equivalent to the $q^2$ defined in the three-body $b\to c$ decay. However, in Fig.~\ref{fig::q2Moment1_fivebody} we plot the $q^2$-moments including a lepton-energy cut which makes the two curves representing the three and five-body decay distinguishable. 

In Figs.~\ref{fig::EMoment_fivebody} -~\ref{fig::q2Moment1_fivebody}, we observe again that the power-correction only give small corrections in the case of lepton energy moments. However, the power-correction are sizable in case of the hadronic invariant mass (Fig.~\ref{fig::MxMoment_fivebody}).


\clearpage

\section{Conclusion}
\label{ch::Conclusion}
The determination of inclusive $V_{cb}$ uses the $\bar{B}\to X_c \ell \bar\nu$ rate, which is obtained from the experimentally measured $\bar{B}\to X \ell$ rate by subtracting (among others) the background signals $\bar{B} \to X_u \ell \bar\nu$ and $\bar{B}\to X_c (\tau \to \ell \bar\nu \nu)\bar\nu$. The goal of this note is to stress that these contributions can be exactly obtained within the local OPE/HQE and thus could be included in an analysis of $B\to X \ell$ without the need to subtract these contributions (which induced uncertainties). 

To facilitate this new strategy, we computed different moments for $b \to u \ell \bar\nu$ at next-to-leading order including power-corrections up to $\mathcal{O}(1/m_b^3)$. We compared our result with generator-level Monte-Carlo data \cite{genmc}. Especially, for hadronic invariant mass moments we note sizable difference between Monte-Carlo and HQE, which could be avoided when using the advocated strategy. 

In addition, we computed for the first time the contributions of $b\to c (\tau \to \ell \bar\nu \nu)\bar\nu$, which contributes at the $4\%$ level. In this case we do not have MC results to compare to, but we present our results in such a way that this comparison could be made in the future. 

In preparation of the Belle II experimental analysis of inclusive $V_{cb}$, which will reach an unprecedented precision, we advocate using the full $B\to X \ell$ rate without subtracting the $b\to u$ and $b\to c(\tau\to \ell \bar\nu \nu) \bar\nu$ contributions. This strategy has the potential to reduce the experimental uncertainties on $V_{cb}$ even further.

\subsubsection*{Acknowledgements}
We thank Danny~van Dyk, Marzia Bordone, Florian~Bernlochner and Lu~Cao for discussions. We specially thank Florian~Bernlochner and Lu~Cao for providing us the generator level Monte-Carlo data. This  research  was supported by the Deutsche Forschungsgemeinschaft (DFG, German Research Foundation)under grant 396021762 - TRR 257.

\appendix

\section{Kinetic mass schemes}
\label{sec::mass-scheme}
The relation between the pole mass scheme $m_{b}(\mu = 0$) and the kinetic scheme $m_{b}^{\text{kin}}(\mu)$ is \cite{Gambino:2007rp, Gambino:2004qm}
\begin{align}
m_{b}(0) = m_{b}^{\text{pole}} &= m_{b}^{\text{kin}}(\mu) + \left[ \bar{\Lambda}(\mu) \right]_{\text{pert}} + \frac{\left[ \mu_{\pi}^{2}(\mu) \right]_{\text{pert}}}{2 m_{b}^{\text{kin}}(\mu)} + \ldots,
\label{eq::m_b-kinetic}
\end{align}
where $\mu$ is the cut-off energy employed in the kinetic mass scheme, which we set $\mu=1$ GeV (see Table~\ref{table::input}) and the  ellipses represent higher-order terms in the $1/m_b$ expansion.  The HQET parameters are now defined as:
\begin{align}
\label{eq::HQE-parameter}
\mu_{\pi}^{2}(0) &= \left(\mu_{\pi}^{2}(\mu)\right)_{\text{kin}} - \left[ \mu_{\pi}^{2}(\mu) \right]_{\text{pert}} , \\
\mu_{G}^{2}(0) &= \left(\mu_{G}^{2}(\mu)\right)_{\text{kin}} - \left[ \mu_{G}^{2}(\mu) \right]_{\text{pert}} , \\
\rho_{LS}^{3}(0) &= \left(\rho_{LS}^{3}(\mu)\right)_{\text{kin}} - \left[ \rho_{LS}^{3}(\mu) \right]_{\text{pert}} , \\
\rho_{D}^{3}(0) &= \left(\rho_{D}^{3}(\mu)\right)_{\text{kin}} - \left[ \rho_{D}^{3}(\mu) \right]_{\text{pert}}.
\end{align}

The quantity $\left[ \bar{\Lambda}(\mu) \right]_{\text{pert}}$ describes the binding energy of the heavy meson and $\left[ \mu_{\pi}^{2}(\mu) \right]_{\text{pert}}$ the residual kinetic energy. Their expression are given as:
\begin{align}
\left[ \bar{\Lambda}(\mu) \right]_{\text{pert}} &= \frac{4}{3} C_{F} \frac{\alpha_{s}(m_{b})}{\pi} \mu \left [ 1 + \frac{\alpha_{s}(m_{b}) \beta_{0} }{2 \pi} \left( \text{log}\left(\frac{m_{b}}{2 \mu} \right) \right) + \frac{8}{3}\right], \\
\left[ \mu_{\pi}^{2}(\mu) \right]_{\text{pert}} &= C_{F} \frac{\alpha_{s}(m_{b})}{\pi} \mu^{2} \left[ 1+ \frac{\alpha_{s}(m_{b}) \beta_{0}}{2\pi} \left(\text{log}\left(\frac{m_{b}}{2 \mu} \right) + \frac{13}{6} \right) \right. \nonumber \\
& \left. - \frac{\alpha_{s}(m_b)}{\pi} C_{A} \left(\frac{\pi^{2}}{6} - \frac{13}{12} \right) \right] + \mathcal{O}\left( \frac{\mu^{3}}{m^3_{b}} \right),   \\
\left[ \rho_{D}^{3}(\mu) \right]_{\text{pert}} &= \frac{2}{3} C_F \frac{\alpha_s(m_b)}{\pi} \mu^3 \left [1+ \frac{\alpha_s(m_b) \beta_0}{2\pi} \left(\log\left(\frac{m_b}{2 \mu}\right) + 2\right) \right. \nonumber \\
& \left. - \frac{\alpha_s(m_b)}{\pi} C_A \left(\frac{\pi^2}{6} -\frac{13}{12} \right) \right] +\mathcal{O}\left( \frac{\mu^{4}}{m^4_{b}} \right) , \\
\left[ \mu_{G}^{2}(\mu) \right]_{\text{pert}} &= \mathcal{O}\left( \frac{\mu^{3}}{m^3_{b}} \right) , \\
\left[ \rho_{LS}^{3}(\mu) \right]_{\text{pert}} &= \mathcal{O}\left( \frac{\mu^{4}}{m^4_{b}} \right).
\end{align}
Here $C_{A} = 3$ and $\beta_{0} = 11 - \frac{2}{3} n_{f}$ with $n_{f} = 3$, i.e. three active massless quarks.

\bibliographystyle{jhep} 
\bibliography{main.bib}

\end{document}